\documentclass[journal,twoside,final]{IEEEtran}

\usepackage{url}
\usepackage{comment}

\usepackage{setspace}
\usepackage{dsfont}
\usepackage{citesort}
\usepackage{cite}
\usepackage{graphicx}
\usepackage{psfrag}
\usepackage{subfigure}
\usepackage{url}
\usepackage{stfloats}
\usepackage{amsmath}
\usepackage{amsfonts}
\usepackage{amssymb}
\usepackage{array}
\usepackage[hypertex]{hyperref}
\usepackage[OT2,OT1]{fontenc}
\newcommand\cyr{%
\renewcommand\rmdefault{wncyr}%
\renewcommand\sfdefault{wncyss}%
\renewcommand\encodingdefault{OT2}%
\normalfont
\selectfont}
\DeclareTextFontCommand{\textcyr}{\cyr}

\usepackage{flushend}
\begin{document}

\title{Generalised Sphere Decoding for \\ Spatial Modulation}

\author{Abdelhamid Younis, Sinan Sinanovi\'c, Marco Di Renzo, Raed Mesleh and Harald Haas
\thanks{%
The associate editor coordinating the review of this paper and approving it for publication was Prof. G. Bauch.
Manuscript received July 29, 2012; revised December 24, 2012.}
\thanks{%
This work was presented in part at the IEEE GLOBECOM 2010, Miami, USA, and IEEE ICC 2011, Kyoto, Japan.}
\thanks{%
A. Younis and H. Haas are with The University of Edinburgh, College of Science and Engineering, Institute for Digital Communications, Joint Research Institute for Signal and Image Processing, King's Buildings, Mayfield Road, Edinburgh, EH9 3JL, UK (e--mail: \{a.younis, h.haas\}@ed.ac.uk).}
\thanks{%
S. Sinanovi\'c is with Glasgow Caledonian University, School of Engineering and Built Environment, George Moore Building, Cowcaddens Road, Glasgow, G4 0BA, UK (e--mail: sinan.sinanovic@gcu.ac.uk).}
\thanks{%
M. Di Renzo is with the Laboratory of Signals and Systems~(L2S), French National Center for Scientific Research~(CNRS),
\'Ecole Sup\'erieure d'\'Electricit\'e~(SUP\'ELEC), University of Paris--Sud XI~(UPS), 3 rue Joliot--Curie, 91192 Gif--sur--Yvette (Paris), France (e--mail: marco.direnzo@lss.supelec.fr).}
\thanks{%
R. Mesleh is with the Electrical Engineering Department and SNCS research center, University of Tabuk, P.O.Box: 71491 Tabuk, Saudi Arabia (e--mail: rmesleh.sncs@ut.edu.sa).}
\thanks{%
Digital Object Identifier 10.1109/TCOMM.2013.09.120547}}

\maketitle

\markboth{IEEE Transactions on Communications, accepted for publication}{Younis \MakeLowercase{\textit{et al.}}: Generalised Sphere Decoding for Spatial Modulation}

\pubid{0090-6778/10\$25.00~\copyright~2013 IEEE}

\pubidadjcol

\begin{abstract}

 In this paper, Sphere Decoding (SD) algorithms for Spatial Modulation (SM) are developed to reduce the computational complexity of Maximum--Likelihood (ML) detectors. Two SDs specifically designed for SM are proposed and analysed in terms of Bit Error Ratio (BER) and computational complexity.
 Using Monte Carlo simulations and mathematical analysis, it is shown that by carefully choosing the initial radius the proposed sphere decoder algorithms offer the same BER as ML detection, with a significant reduction in the computational complexity.
 A tight closed form expression for the BER performance of SM--SD is derived in the paper, along with an algorithm for choosing the initial radius which provides near to optimum performance.  Also, it is shown that none of the proposed SDs are always superior to the others, but the best SD to use depends on the target spectral efficiency. The computational complexity trade--off offered by the proposed solutions is studied via analysis and simulation, and is shown to validate our findings. Finally, the performance of SM--SDs are compared to Spatial Multiplexing (SMX) applying ML decoder and applying SD.
 It is shown that for the same spectral efficiency, SM--SD offers up to~$84\%$ reduction in complexity compared to SMX--SD, with up to~$1$ dB better BER performance than SMX--ML decoder.
\end{abstract}


\section{Introduction} \label{sec1}

 \PARstart{M}{ultiple}--input multiple--output (MIMO) systems offer a significant increase in spectral efficiency in comparison to single antenna systems~\cite{t9902}. An example is Spatial Multiplexing (SMX)~\cite{f9601}, which transmits simultaneously over all the transmit antennas. This method achieves a spectral efficiency that increases linearly with the number of transmit antennas.
 However, these systems cannot cope with the exponential increase of wireless data traffic, and a larger number of transmit antennas (large scale MIMO) should be used~\cite{hbd1101}. Large scale MIMO studied in~\cite{mzcr0901,cv0901,mcs0801}, offers higher data rates and better bit error rate (BER) performance.
 However, this comes at the expense of an increase in:
\begin{enumerate}
 \item  Computational complexity: A SMX maximum likelihood (ML) optimum receiver searches across all possible combinations, and tries to resolve the inter--channel interference (ICI) caused by transmitting from all antennas simultaneously, on the same frequency. Sphere decoder (SD) was proposed to reduce the complexity of the SMX--ML algorithm while retaining a near optimum performance~\cite{vb9901,dcb0001}. The SD reduces the complexity of the ML decoder by limiting the number of possible combinations. Only those combinations that lie within a sphere centred at the received signal are considered.
 However, even though SMX--SD offers a large reduction in complexity compared to SMX--ML, it still has a high complexity which increases with the increase of the number of transmit antennas.
 \item Hardware complexity: In SMX the number of radio frequency (RF) chains is equal to the number of transmit antennas. From~\cite{mw0403}, RF chains are circuits that do not follow Moore's law in progressive improvement. Therefore, increasing the number of transmit antennas and consequently the number of RF chains increases significantly the cost of real system implementation~\cite{ssrhg1201}.
 \item Energy consumption: RF chains contain Power Amplifiers (PAs) which are responsible for $50$--$80\%$ of the total power consumption in the transmitter~\cite{czbfkgav1011}. Therefore, increasing the number of RF chains results in a decrease in the energy efficiency~\cite{ssrhg1201}.
 \end{enumerate}

\noindent Thus, SMX may not always be feasible and a more energy efficient and low complexity solution should be considered.\pubidadjcol

 Spatial Modulation (SM) is a transmission technology proposed for MIMO wireless systems. It aims to increase the spectral efficiency, $(m)$, of single--antenna systems while avoiding ICI~\cite{mhsay0801}. This is attained through the adoption of a new modulation and coding scheme, which foresees: i)~the activation, at each time instance, of a single antenna that transmits a given data symbol (\emph{constellation symbol}), and ii)~the exploitation of the spatial position (index) of the active antenna as an additional dimension for data transmission (\emph{spatial symbol})~\cite{rhg1101}. Both the \emph{constellation symbol} and the \emph{spatial symbol} depend on the incoming data bits. An overall increase by the base--two logarithm of the number of transmit--antennas of the spectral efficiency is achieved. This limits the number of transmit antennas to be a power of two unless fractional bit encoding SM (FBE--SM)~\cite{srsmh1001}, or generalised SM (GSM)~\cite{ysmh1001} are used. Activating only one antenna at a time means that only one RF chain is needed, which significantly reduces the hardware complexity of the system~\cite{jgsc0901}. Moreover, as only one RF chain is needed, SM offers a reduction in the energy consumption which scales linearly with the number of transmit antennas~\cite{ssrh1201,ssrhg1201}. Furthermore, the computational complexity of SM--ML is equal to the complexity of single--input multiple--output (SIMO) systems~\cite{jgs0801}, \emph{i.e.} the complexity of SM--ML depends only on the spectral efficiency and the number of receive antennas, and does not depend on the number of transmit antennas. Recently the potential benefits of SM have been validated not only by simulations but also via experiments~\cite{symcrwgbh01,syh1201}. Moreover, in~\cite{ytrwbhg1301} for the first time the performance of SM is analysed using real--world channel measurements. Accordingly, SM appears to be a good candidate for large scale MIMO~\cite{dh1201a,dh1301a,rlgh1201,rh1102}.

 In spite of its low--complexity implementation, there is still potential for further reductions, by limiting the number of possible combinations using the SD principle. However, existing SD algorithms in literature do not consider the basic and fundamental principle of SM,  that only one antenna is active at any given time instance. Therefore, two modified SD algorithms based on the tree search structure that are tailored to SM are proposed. The first SD will be called receiver--centric SD (SM--Rx), which was first presented in~\cite{ymhg1001}. The algorithm in~\cite{ymhg1001} combines the received signal from the multiple receive antennas, as long as the Euclidean distance from the received point is less than a given radius. This SD--based detector is especially suitable when the number of receive--antennas is very large. This technique reduces the size of the search space related to the multiple antennas at the receiver (we denote this search space as ``receive search space''). It will be shown later that there is no loss in either the diversity order or the coding gain, \emph{i.e.} the BER is very close to that of the ML detector. However, the main limitation is that it does not reduce the search space related to the number of possible transmitted points (we denote this as ``transmit search space''). This limitation prevents the detector from achieving a significant reduction in computational complexity when high data rates are required.

 The second SD, which is called Transmit--centric (SM--Tx) was first presented in~\cite{yrmh1101}. It aims at reducing the transmit search space by limiting the number of possible spatial and constellation points. The SM--Tx algorithm avoids an exhaustive search by examining only those points that lie inside a sphere with a given radius. However, SM--Tx is limited to the overdetermined MIMO setup $\left(N_t\leq N_r\right)$, where $N_t$ and $N_r$ are the number of transit and receiver antennas respectively.
 In~\cite{rh1201_gen,rh1201a}, it is shown that SM--Tx in~\cite{yrmh1101} can still be used for the case of $(2N_r-1)\geq N_t>N_r$, where SM--Tx is referred to as enhanced Tx--SD (E--Tx--SD).
 Moreover, in~\cite{rh1201_gen,rh1201a} a detector for the case of $N_t>N_r$ referred to as generalised Tx--SD (G--Tx--SD) is proposed.
  By using the division algorithm the G--Tx--SD technique:
  1) Divides the set of possible antennas to a number of subsets.
  2) Performs E--Tx--SD over each subset.
  3) Takes the minimum solution of all the sets.
 However, in this paper we propose a simple solution, in which all that is needed is to set a constant $\varphi$ to $0$ for $N_t\leq N_r$ and $\varphi=\sigma_n^2$ for $N_t>N_r$, where $\sigma_n^2$ is the noise variance.
 In~\cite{rh1201_gen,rh1201a}, the normalised expected number of nodes visited by the SM--Tx algorithm is used to compare its complexity with the complexity of the SM--ML algorithm. This does not take into account pre--computations needed by SM--Tx. In this paper, when comparing the complexity of SM--Tx with the complexity of SM--ML and SM--Rx, we take into account the pre--computations needed by the SM--Tx. We show that because of those pre--computations, the SM--Tx technique is not always the best solution, where in some cases it is even more complex than SM--ML.
 The performance of both SDs is closely tied to the choice of the initial radius. The chosen radius should be large enough for the sphere to contain the solution. On the one hand, the larger the radius is, the larger the search space, which increases the complexity. On the other hand, a small radius may cause the algorithm to fail in finding a point inside the sphere.

 In this paper, a careful study of the performance of these two detectors is provided, along with an accurate comparison of their computational complexity. Numerical results show that with no loss in the BER performance, the proposed solutions provide a substantial reduction in computational complexity with respect to the SM--ML decoder. We also derive a closed form expression for the BER performance of SM--SD and show that the initial radius can be chosen such that SM--SD gives an optimum performance. Furthermore, it is shown that SM--Rx is less complex than SM--Tx for lower spectral efficiencies, while SM--Tx is the best solution for higher spectral efficiencies. Finally, using numerical results we show that SM--SD offers a significant reduction and nearly the same performance when compared to SMX with ML decoder or SD.

 The remainder of this paper is organised as follows: In section \ref{sec:SysModel}, the system model along with the ML--optimum detector is summarised. In section \ref{sec:SD}, the new SM--Rx and SM--Tx receivers are described. In section \ref{sec:Comp}, an accurate analysis of the computational complexity of both SM--Rx and SM--Tx is performed. In section \ref{sec:ANA}, the analytical BER performance for SM--SDs is derived along with the initial radius selection method. Numerical results are presented in section \ref{sec:Result}, and the paper is concluded in section \ref{sec:Con}.

\section{System Model} \label{sec:SysModel}

\subsection{SM Modulator}
 In SM, the bit stream emitted by a binary source is divided into blocks containing $m=\log_2\left(N_t\right) + \log_2\left(M\right)$ bits each, where $M$ is the constellation size. Then the following mapping rule is used~\cite{mhsay0801}:

\begin{itemize}
 \item The first $\log_2\left(N_t\right)$ bits are used to select the antenna which is switched on for data transmission, while the other transmit antennas are kept silent. In this paper, the actual transmit antenna which is active for transmission is denoted by $\ell_t$ with $\ell_t \in \{1,2,\dots,N_t\}$.
 \item The second $\log_2\left(M\right)$ bits are used to choose a symbol in the signal--constellation diagram. Without loss of generality, Quadrature Amplitude Modulation (QAM) is considered. In this paper, the actual complex symbol emitted by the transmit antenna $\ell_t$ is denoted by $s_t$, with $s_t \in \{s_1,s_2,\dots,s_{M}\} $.
\end{itemize}


 Accordingly, the $N_t \times 1 $ transmitted vector is:
\begin{equation}
{\bf{x}}_{\ell_t ,s_t }  = \left[ {{\bf{0}}_{1 \times \left( {\ell_t  - 1} \right)} ,s_t ,{\bf{0}}_{1 \times \left( {N_t  - \ell_t } \right)} } \right]^T, \label{Eq:Mod_vec}
\end{equation}

\noindent where $\left[  \cdot  \right]^T$ denotes transpose operation, and ${\bf{0}}_{p \times q}$ is a $p \times q$ matrix with all--zero entries. Note, a power constraint on the average energy per transmission of unity is assumed (\emph{i.e.} $E_s = \mathrm{E}[\mathbf{x}^{H}\mathbf{x}]=1$), where $\mathrm{E}\{\cdot\}$ is the expectation operator.

 From above, the maximum achievable spectral efficiency by SM is,

\begin{equation} \vspace{0.2cm}
m_{\text{SM}} = \log_2(N_t) + \log_2(M) \label{eq1}
\end{equation}

However, for SMX,

\begin{equation}\vspace{0.2cm}
 m_{\text{SMX}} = N_t \log_2(M) \label{eq2}
\end{equation}

From \eqref{eq1} and \eqref{eq2}, we can see that the spectral efficiency of SM does not increase linearly with the number of transmit antennas as SMX does. Therefore, SM needs a larger number of transmit antennas/ larger constellation size to arrive at the same spectral efficiency as SMX. However, because in SM only one antenna is active:

\begin{itemize}
 \item The computational complexity of SM does not depend on the number of transmit antennas. Unlike SMX where the computational complexity increases linearly with the number of transmit antennas, the computational complexity of SM is the same as the computational complexity of SIMO systems.
 \item The number of RF chains needed by SM is significantly less than the number of RF chains needed by SMX. In fact, only one RF chain is required for SM.
\end{itemize}

For these reasons we believe that SM is a strong candidate for large scale MIMO systems, which strongly motivates this work.

\subsection{Channel Model}

The modulated vector, ${\bf{x}}_{\ell_t ,s_t }$, in \eqref{Eq:Mod_vec} is transmitted through a frequency--flat $N_r\times N_t$ MIMO fading channel with transfer function $\mathbf{H}$, where $N_r$ is the number of receive antennas. In this paper, a Rayleigh fading channel model is assumed. Thus, the entries of $\mathbf{H}$ are modelled as complex independent and identically distributed (i.i.d.) entries according to $\mathcal{CN}(0, 1)$. Moreover, a perfect channel state information (CSI) at the receiver is assumed, with no CSI at the transmitter.

 Thus, the $N_r \times 1$ received vector can be written as follows:

\begin{eqnarray}
{\bf{y}} &=& {\bf{Hx}}_{\ell_t ,s_t }  + {\bf{n }} \nonumber \\
         &=& {\bf{h}}_{\ell_t}s_t + \bf{n} \label{Eq:Comp_Mod}
\end{eqnarray}

\noindent where ${\bf{n }}$ is the $N_r$--dimensional Additive White Gaussian Noise (AWGN) with zero--mean and variance $\sigma^2$ per dimension at the receiver input, and ${\bf{h}}_{\ell }$ is the $\ell$--th column of $\mathbf{H}$. Note, the signal-to-noise-ratio is $\text{SNR} = E_s/N_o = 1/\sigma_n^2$.

\subsection{ML--Optimum Detector}

The ML optimum receiver for MIMO systems can be written as,

\begin{equation}
 \hat{\mathbf{x}}_t^{\left(\text{ML}\right)} =  \mathop {\arg \min }\limits_{\scriptstyle \mathbf{x} \in \mathcal{Q}^m} \left\{ {\left\| {{\bf{y}} - {\bf{H}}\mathbf{x}} \right\|_{\rm{F}}^2 } \right\} \label{Eq:MLMIMO}
\end{equation}

\noindent where $\mathcal{Q}^m$ is a $2^m$ space containing all possible $\left(N_t\times 1\right)$ transmitted vectors, $\left\|  \cdot  \right\|_{\rm{F}}$ is the Frobenius norm, and ${\hat  \cdot }$ denotes the estimated spatial and constellation symbols.

Note, in SM only one transmit antenna is active at a time. Therefore, the optimal receiver in \eqref{Eq:MLMIMO} can be simplified to,
\begin{eqnarray}\vspace{0.2cm}
 \left[ {\hat \ell_t^{\left( {{\rm{ML}}} \right)} ,\hat s_t^{\left( {{\rm{ML}}} \right)} } \right] &=& \mathop {\arg \min }\limits_{\scriptstyle \ell \in \left\{ {1,2, \ldots N_t } \right\} \hfill \atop
  \scriptstyle s \in \left\{ {s_1 ,s_2 , \ldots s_M } \right\} \hfill} \left\{ {\left\| {{\bf{y}} - {\bf{h}}_\ell s} \right\|_{\rm{F}}^2 } \right\} \nonumber \\
  &=& \mathop {\arg \min }\limits_{\scriptstyle \ell \in \left\{ {1,2, \ldots N_t } \right\} \hfill \atop
  \scriptstyle s \in \left\{ {s_1 ,s_2 , \ldots s_M } \right\} \hfill} \left\{ {\sum\limits_{r = 1}^{N_r } {\left| {y_r  - h_{\ell,r} s} \right|^2 } } \right\} \nonumber  \\ \label{Eq:ML}
\end{eqnarray}

\noindent where $y_r$ and $h_{\ell,r}$ are the $r$--th entries of ${\bf{y}}$ and ${\bf{h}}_{\ell}$ respectively.

\section{Sphere Decoders for SM} \label{sec:SD}

In this section we introduce two SDs tailored for SM, SM--Rx and SM--Tx. SM--Rx aims at reducing the number of summations over $N_r$ required by the ML receiver in \eqref{Eq:ML}. SM--Tx aims at reducing the number of points $(\ell,s)$ the ML receiver searches over.

 First, for ease of derivation, we introduce the real--valued equivalent of the complex--valued model in \eqref{Eq:Comp_Mod} following~\cite{book:th01},

\begin{eqnarray}
{\bf{\bar y}} &=& {\bf{\bar H\bar x}}_{\ell_t ,s_t }  + {\bf{\bar n}} \nonumber \\
              &=& {\bf{\bar h}}_{\ell_t}{\bf{\bar s}}_t + \bf{\bar n}
\end{eqnarray}

\noindent where,

\begin{eqnarray}
 {\bf{\bar y}} &=& \left[ {{\mathop{\rm Re}\nolimits} \left\{ {{\bf{y}}^T } \right\},{\mathop{\rm Im}\nolimits} \left\{ {{\bf{y}}^T } \right\}} \right]^T   \\
 {\bf{\bar H}} &=& \left[ {\begin{array}{*{20}c}
   {{\mathop{\rm Re}\nolimits} \left\{ {\bf{H}} \right\}} & {{\mathop{\rm Im}\nolimits} \left\{ {\bf{H}} \right\}}  \\
   { - {\mathop{\rm Im}\nolimits} \left\{ {\bf{H}} \right\}} & {{\mathop{\rm Re}\nolimits} \left\{ {\bf{H}} \right\}}  \\
\end{array}} \right]  \\
  {\bf{\bar x}}_{\ell ,s }  &=& \left[ {{\mathop{\rm Re}\nolimits} \left\{ {{\bf{x}}_{\ell ,s }^T } \right\},{\mathop{\rm Im}\nolimits} \left\{ {{\bf{x}}_{\ell ,s }^T } \right\}} \right]^T \label{Eq_9}  \\
 {\bf{\bar n}} &=& \left[ {{\mathop{\rm Re}\nolimits} \left\{ {{\bf{n}}^T } \right\},{\mathop{\rm Im}\nolimits} \left\{ {{\bf{n}}^T } \right\}} \right]^T  \\
 \bar{\bf h}_{\ell} &=& \left[ \bar{\bf H}_{\ell} , \bar{\bf H}_{\ell+N_t} \right] \\
\bar{\bf s} &=& \left[ {\begin{array}{*{20}c}
   \mathop{\rm Re}\{s\} \\
   \mathop{\rm Im}\{s\} \\
\end{array}} \right]
\end{eqnarray}

\noindent where ${\mathop{\rm Re}\nolimits} \left\{  \cdot  \right\}$ and ${\mathop{\rm Im}\nolimits} \left\{  \cdot  \right\}$ denote real and imaginary parts respectively, and ${\bar{\bf H}}_{\ell}$ is the $\ell$--th column of $\bf{\bar H}$.

 \begin{figure*}[ht]
\setcounter{equation}{21}

\begin{equation}
 R_{i+1}^2 = {\left\| {{\bf{\bar z}} - {\bf{\bar D\bar x}}_{\ell,s} } \right\|_{\rm{F}}^2 }
 = (R_i^2-R'^2) + \sum_{\nu=1}^{N_t} \left( z_{\nu} - D_{\left(\nu,\ell\right)}{\rm Re}\left\lbrace s \right\rbrace - D_{\left(\nu,\ell+N_t\right)}{\rm Im}\left\lbrace s \right\rbrace \right)^2 \label{Eq_R_new}
\end{equation}
\normalsize \hrulefill \vspace*{0pt}
\setcounter{equation}{13}
\end{figure*}

\subsection{SM--Rx Detector}\label{Rx-SD}

 The SM--Rx is a reduced--complexity and close--to--optimum BER--achieving decoder, which aims at reducing the receive search space. The detector can formally be written as follows:

\begin{equation}
\left[ {\hat \ell_t^{\left( {{\rm{Rx}}} \right)} ,\hat s_t^{\left( {{\rm{Rx}}} \right)} } \right] = \mathop {\arg \min
}\limits_{\scriptstyle \ell \in \left\{ {1,2, \ldots N_t } \right\} \hfill \atop{
  \scriptstyle s \in \left\{ {s_1 ,s_2 , \ldots s_M } \right\} \hfill}}  \left\{ {\sum\limits_{r = 1}^{\tilde N_r \left( {\ell,s} \right)} {\left| {\bar{y}_r  - \bar{\bf{h}}_{\ell,r} \bf{\bar{s}}} \right|^2 } } \right\} \label{Eq:SM--Rx}
\end{equation}

\noindent where $\bar{\bf{h}}_{\ell,r}$ is the $r$--th row of $\bar{\bf{h}}_{\ell}$, and,
\begin{equation}
\tilde N_r \left( {\ell,s} \right) = \mathop {\max }\limits_{n \in \left\{ {1,2, \ldots 2N_r } \right\}} \left\{ { n
\left| \sum\limits_{r = 1}^n {\left| {y_r  - h_{\ell,r} s} \right|^2 }  \leq R^2 \right. } \right\} \label{Eq:NrTilde}
\end{equation}

 The idea behind SM--Rx is that it keeps combining the received signals as long as the Euclidean distance in \eqref{Eq:SM--Rx} is less or equal to the radius $R$. Whenever a point is found to be inside the sphere, the radius, $R$, is updated with the Euclidean distance of that point. The point with the minimum Euclidean distance and $\tilde N_r\left(\cdot,\cdot\right)=2N_r$ is considered to be the solution.

\subsection{SM--Tx Detector} \label{Tx-SD}

 The conventional SD is designed for SMX, where all antennas are active at each time instance~\cite{hv0501a,vb9901,ct0501,wl0901}. However, in SM only one antenna is active at a time. Therefore, a modified SD algorithm tailored for SM, named SM--Tx, is presented. More specifically, similar to conventional SDs, the SM--Tx scheme reduces the number of points $\left( {\ell,s} \right)$ for $\ell \in \left\{ {1,2, \ldots N_t } \right\}$ and $s \in \left\{ {s_1 ,s_2 , \ldots s_M } \right\}$ to be searched through in \eqref{Eq:ML}, \emph{i.e.}, the transmit search space, by computing the Euclidean distances only for those points that lie inside a sphere with radius $R$ and are centred around the received signal. However, unlike conventional SDs, in our scheme the set of points inside the sphere are much simpler to compute, as there is only a single active antenna in SM. In this section, we show how to compute the set of points.

 The Cholesky factorisation of the $(2N_t \times 2N_t)$ positive definite matrix ${\bf{\bar{G}}}={\bf{\bar{H}}}^T{\bf{\bar{H}}}+\varphi{\bf{\bar{I}}}_{N_t}$ is $\mathbf{\bar G} = \mathbf{\bar D}\mathbf{\bar D}^{T}$, where

 \begin{equation}
  \varphi = \left\{ \begin{array}{rl}
                    \sigma_n^2 & N_t>N_r \\
		    0          & N_t\leq N_r \\
                   \end{array} \right.
 \end{equation}

 Then the SM--Tx can be formally written as follow,

 \begin{equation}
 \left[ {\hat \ell_t^{\left( {{\rm{Tx - SD}}} \right)} ,\hat s_t^{\left( {{\rm{Tx - SD}}} \right)} } \right] = \mathop {\arg \min }\limits_{\left( {\ell,s} \right) \in \Theta _R } \left\{ {\left\| {{\bf{\bar z}} - {\bf{\bar D\bar x}}_{\ell,s} } \right\|_{\rm{F}}^2 } \right\} \label{Eq:Tx-SD}
\end{equation}

\noindent where ${\Theta _R }$ is the subset of points $\left( {\ell,s} \right)$ for $\ell \in \left\{ {1,2, \ldots N_t } \right\}$ and $s \in \left\{ {s_1 ,s_2 , \ldots s_M } \right\}$ in the transmit search space that lie inside a sphere with radius $R$ and centred around the received signal ${{\bf{\bar z}}}$, $\bar{\bf z}=\bar{\bf D}\bar{\rho}$ and $\bar{\rho}=\mathbf{\bar{G}}^{-1}\mathbf{\bar{H}}^T\mathbf{\bar{y}}$.

 Unlike SM--Rx, SM--Tx reduces the computational complexity of the ML receiver by reducing the transmit search space, which is done by the efficient computation of the subset ${\Theta _R }$.
 After some algebraic manipulations as shown in Appendix \ref{App:Int}, the subset of points ${\Theta _R }$ lie in the intervals:

\begin{equation}
\label{Eq:int1}   \frac{{ - R_i + \bar z_{\ell+N_t} }}{{\bar D_{(\ell+N_t,\ell+N_t)} }} \le {\rm Im}\left\lbrace s \right\rbrace  \le \frac{{R_i  + \bar z_{\ell+N_t} }}{{\bar D_{(\ell+N_t,\ell+N_t)} }}
\end{equation}

\begin{equation}
\label{Eq:int2} \frac{{ - R'  + \bar z_{\ell|\ell + N_t } }}{{\bar D_{\ell,\ell} }} \le {\rm Re}\left\lbrace s \right\rbrace \le \frac{{R'  + \bar z_{\ell|\ell + N_t} }}{{\bar D_{\ell,\ell} }}
\end{equation}

\noindent where
\begin{eqnarray}
\bar z_{a|b }  &=& \bar z_{a}  - \bar D_{\left(a,b\right) }{\rm Im}\left\lbrace s \right\rbrace \\
R'^2 &=& R^2 - \sum_{\nu=N_t+1}^{2N_t} \bar z^2_{\nu|\ell+N_t} \label{Eq_R_dash}
\end{eqnarray}

Note, every time a point is found inside the sphere, the radius $R$ is updated as shown in \eqref{Eq_R_new}, with the Euclidean distance of that point. Moreover, (\ref{Eq:int2}) needs to be computed only for those points that lie inside the interval in (\ref{Eq:int1}), for the reason that (\ref{Eq:int2}) depends implicitly on (\ref{Eq:int1}).

\setcounter{equation}{22}

Because of the unique properties of SM the intervals in (\ref{Eq:int1}) and (\ref{Eq:int2}) needs to be calculated only once for each possible transmit point, unlike conventional SDs where the intervals have to be calculate $N_t$ times for each transmit point. Furthermore, we note that SM--Tx works for both underdetermined MIMO setup with $N_t>N_r$, and overdetermined MIMO setup with $N_t\leq N_r$.

 As opposed to the SM--Rx scheme, the SM--Tx scheme uses some pre--computations to estimate the points that lie inside the sphere of radius $R$. These additional computations are carefully taken into account in the analysis of the computational complexity of the SM--Tx scheme and its comparison with the ML--optimum detector in section \ref{sec:Comp}.

 \begin{figure*}[hb]
\normalsize \hrulefill \vspace*{0pt}
\setcounter{equation}{33}
\begin{equation}
 \Pr_{e,\text{SM--SD}} \leq \left( \Pr\left((\tilde{\ell}_{\text{SM--ML}},\tilde{s}_{\text{SM--ML}})\neq \left(\ell_t,s_t\right)\right) + \Pr\left(\left(\ell_t,s_t\right) \notin \Theta _R\right) \right) \label{Eq_44}
\end{equation}
\setcounter{equation}{22}
\end{figure*}

\section{Computational Complexity of SM--Rx and SM--Tx} \label{sec:Comp}

In this section, we analyse the computational complexity of SM--ML, SM--Rx and SM--Tx. The complexity is computed as the number of real multiplication and division operations needed by each algorithm~\cite{book:gl9601}.

\subsection{SM--ML}

The computational complexity of SM--ML receiver in \eqref{Eq:ML}, yields,

\begin{equation}
 \mathcal{C}_{\text{SM--ML}} = 8N_r2^m, \label{Eq:MLcomp}
\end{equation}

\noindent as the ML detector searches through the whole transmit and receive search spaces. Note, evaluating the Euclidean distance $\left({\left| {y_r  - h_{\ell,r} s} \right|^2 }\right)$ requires 8 real multiplications.

The computational complexity of SMX--ML receiver in \eqref{Eq:MLMIMO} is equal to

\begin{equation}
 \mathcal{C}_{\text{SMX--ML}} = 4\left(N_t+1\right) N_r 2^m. \label{Eq:CompSMX}
\end{equation}

\noindent Note, $\left(\left|\mathbf{y}-\mathbf{Hx}\right|^2\right)$ in \eqref{Eq:MLMIMO} requires $\left(N_t+1\right)$ complex multiplications.

From \eqref{Eq:MLcomp} and \eqref{Eq:CompSMX}, the complexity of SM does not depend on the number of transmit antennas, and it is equal to the complexity of SIMO systems. However, the complexity of SMX increases linearly with the number of transmit antennas.

 Thus, the reduction of SM--ML receiver complexity relative to the complexity of the SMX--ML decoder for the same spectral efficiency is given by,

 \begin{equation}
  \mathcal{C}^{\text{ML}}_{\text{rel}} = 100 \times \left( 1 - \frac{2}{N_t+1} \right). \label{Eq:RelComp}
 \end{equation}

\noindent From \eqref{Eq:RelComp}, the reduction in complexity offered by SM increases with the increase in the number of transmit antennas. For example for $N_t=4$ SM offers a $60\%$ reduction in complexity compared to SMX, and as the number of transmit antennas increases the reduction increases.

\subsection{SM--Rx}

The complexity of the SM--Rx receiver is given by:

\begin{equation}
\label{Eq:Rxcomp} \mathcal{C}_{{\rm{Rx-SD}}} = 3\sum\limits_{\ell = 1}^{N_t } {\sum\limits_{s = 1}^M {\tilde N_r \left( {\ell,s} \right)} }
\end{equation}

It is easy to show that $\mathcal{C}_{{\rm{Rx-SD}}}$ lies in the interval $3 \times 2^m  \le \mathcal{C}_{{\rm{Rx-SD}}} \le 6N_r2^m$, where the lower
bound corresponds to the scenario where $\tilde N_r \left( {\ell,s} \right) = 1$, and the upper bound corresponds to the scenario where $\tilde N_r \left( {\ell,s} \right) = 2N_r$ for $\ell \in \left\{ {1,2, \ldots N_t } \right\}$ and $s \in \left\{ {s_1 ,s_2 , \ldots s_M } \right\}$.
 An interesting observation is that SM--Rx offers a reduction in complexity even for the case of $N_r=1$, where the complexity lies in the interval $3 \times 2^m \leq \mathcal{C}_{\rm{Rx-SD}} \leq 6 \times 2^m$.
We note that the SM--Rx solution requires no pre--computations with respect to the ML--optimum detector. In fact, $\tilde N_r \left( {\ell,s} \right)$ for $\ell \in \left\{ {1,2, \ldots N_t } \right\}$ and $s \in \left\{ {s_1 ,s_2 , \ldots s_M } \right\}$ in \eqref{Eq:NrTilde} are implicitly computed when solving the detection problem in \eqref{Eq:SM--Rx}.

\subsection{SM--Tx}
The computational complexity of SM--Tx can be upper--bounded by,

\begin{equation}
 \label{Eq:Txcomp} \mathcal{C}_{{\rm{Tx - SD}}} \leq \mathcal{C}_{\Theta_R }  + 3N_t{\rm{card}}\left\{ {\Theta _R } \right\}
\end{equation}

\noindent where $\rm{card\{\cdot\}}$ denotes the cardinality of a set, and $\mathcal{C}_{\Theta_R}$ is the complexity of finding the points in the subset $\Theta_R$,

\begin{equation}
 \mathcal{C}_{\Theta_R} = \mathcal{C}_{\text{Pre-Comp}} + \mathcal{C}_{\text{Interval}}
\end{equation}

\noindent where,

\begin{enumerate}

\item $\mathcal{C}_{\text{Pre-Comp}}$ is the number of operations needed to compute the Cholesky decomposition. Calculating the upper triangular matrix $\mathbf{\bar D}$ using Cholesky decomposition has the complexity~\cite{book:gl9601},

\begin{equation}
 \mathcal{C}_{\rm CH} = 4N_t^3/3
\end{equation}

 \noindent It can be easily shown that the calculation of $\bar{\bf G}$, $\bar \rho$ and $\mathbf{\bar z}$ requires $2N_rN_t(2N_t+1)$, $2N_t(2N_t+2N_r+1)$ and $N_t(2N_t+1)$ real operations respectively, where back--substitution algorithm was used for calculating $\bar \rho$~\cite{book:gl9601}.

 \noindent Hence,
\begin{equation}
 \mathcal{C}_{\text{Pre-Comp}} = \mathcal{C}_{\rm CH} + N_t(4N_rN_t+6N_r+6N_t+3)
\end{equation}

\item $\mathcal{C}_{\text{Interval}}$ is the number of operations needed to compute the intervals in \eqref{Eq:int1},\eqref{Eq:int2},

\begin{equation}
 \mathcal{C}_{{\rm{interval}}} = 2N_t  + (2N_t+3)N_{\eqref{Eq:int2}}
\end{equation}

\noindent where,

\begin{itemize}
 \item For \eqref{Eq:int1}: $2N_t$ real divisions are needed.
 \item For \eqref{Eq:int2}: $\left(2N_t+3\right)N_{\eqref{Eq:int2}}$ real multiplications are needed, where $\left(2N_t+3\right)$ is the number of real computations needed to compute \eqref{Eq:int2}, and  $N_{\eqref{Eq:int2}}$ is the number of times \eqref{Eq:int2} is computed, which is calculated by simulations.
 Note, i) the interval in \eqref{Eq:int2} depends on the antenna index $\ell$ and only the imaginary part of the symbol $s$, ii) some symbols share the same imaginary part. Therefore, \eqref{Eq:int2} is only  calculated for those points $(\ell,s)$ which lie in the interval in \eqref{Eq:int1} and does not have the same $\ell$ and $\mathop{\rm Im}\{s\}$ as a previously calculated point.
\end{itemize}

\end{enumerate}

 \begin{figure*}[ht]
\setcounter{equation}{45}

\begin{equation}
\mathrm{E}_{\bf{H}}\left\lbrace \Pr_{e,\text{SM--SD}} \right\rbrace = \left[ \zeta\left(\frac{\sigma_s^2}{4\sigma_n^2}\right) \right]^{N_r} \sum_{r=0}^{N_r-1}\binom{N_r-1+r}{r}\left[1-\zeta\left(\frac{\sigma_s^2}{4\sigma_n^2}\right)\right]^r \label{Eq:EFinal}
\end{equation}
\normalsize \hrulefill \vspace*{0pt}
\setcounter{equation}{31}
\end{figure*}

\section{Error Probability of SM--SDs and Initial Radius Selection Method} \label{sec:ANA}

 In this section, we derive an analytical expresion for the BER performance of SM--SD, and we show that SM--SD offers a near optimum performance. The BER for SM--SD is estimated using the union bound~\cite{book:p0001}, which can be expressed as follows,

\begin{equation}
 \underset{\text{SM--SD}}{\text{BER}} \leq \sum_{\ell_t,s_t}\sum_{\ell,s}\frac{N\left(\mathbf{\bar x}_{\ell_t,s_t},\mathbf{\bar x}_{\ell,s}\right)}{m}\frac{\mathrm{E}_{\bf{H}}\left\lbrace \Pr_{e,\text{SM--SD}} \right\rbrace }{2^m} \label{Eq:BER_1}
\end{equation}

\noindent where $N\left(\mathbf{\bar x}_{\ell_t,s_t},\mathbf{\bar x}_{\ell,s}\right)$ is the number of bits in error between $\mathbf{\bar x}_{\ell_t,s_t}$ and $\mathbf{\bar x}_{\ell,s}$, and,

\begin{equation}
 \Pr_{e,\text{SM--SD}} = \Pr\left(\left(\tilde{\ell}_{\text{SM--SD}},\tilde{s}_{\text{SM--SD}}\right)\neq \left(\ell_t,s_t\right)\right)
\end{equation}

\noindent is the pairwise error probability of deciding on the point $\left(\tilde{\ell}_{\text{SM--SD}},\tilde{s}_{\text{SM--SD}}\right)$ given that the point $\left(\ell_t,s_t\right)$ is transmitted.

 The probability of error $\Pr_{e,\text{SM--SD}}$ can be thought of as two mutually exclusive events depending on whether the transmitted point $\left(\ell_t,s_t\right)$ is inside the sphere. In other words, the probability of error for SM--SD can be separated in two parts, as shown in \eqref{Eq_44}~\cite{jbot0901}:

\begin{itemize}
 \item $\Pr\left((\tilde{\ell}_{\text{SM--ML}},\tilde{s}_{\text{SM--ML}})\neq \left(\ell_t,s_t\right)\right)$: The probability of deciding on the incorrect transmitted symbol and/or used antenna combination, given that the transmitted point $\left(\ell_t,s_t\right)$ is inside the sphere.
 \item $\Pr\left(\left(\ell_t,s_t\right) \notin \Theta _R\right)$: The probability that the transmitted point $\left(\ell_t,s_t\right)$ is outside the set of points $\Theta _R$ considered by the SD algorithm.
\end{itemize}

\setcounter{equation}{34}

 However, the probability of error for the ML decoder is,
\begin{equation}
 \Pr_{e,\text{SM--ML}} \leq \Pr\left((\tilde{\ell},\tilde{s})\neq \left(\ell_t,s_t\right)\right) \label{Eq_45}
\end{equation}

\noindent Thus, SM--SD will have a near optimum performance when,
 \begin{equation}
 \Pr\left(\left(\ell_t,s_t\right) \notin \Theta _R\right) << \Pr\left((\tilde{\ell},\tilde{s})\neq \left(\ell_t,s_t\right)\right) \label{Eq:NearOP}
\end{equation}

  The probability of \emph{not} having the transmitted point $\left(\ell_t,s_t\right)$ inside $\Theta_R$ can be written as,
\begin{eqnarray}
 \Pr\left(\left(\ell_t,s_t\right) \notin \Theta _R\right) &=& \Pr\left(\sum_{r=1}^{2N_r}\left| {\bar{y}_r  - \bar{\bf{h}}_{\ell_t,r} \bar{\bf{s}}_t} \right|^2>R^2\right) \nonumber \\
   &=& \Pr\left(\kappa > \left(\frac{R}{\sigma_n/\sqrt{2}}\right)^2\right) \nonumber \\
&=& 1 - \frac{\gamma\left(N_r,\left(\frac{R}{\sigma_n}\right)^2\right)}{\Gamma(N_r)} \label{Eq:PrNoPoint}
\end{eqnarray}
\noindent where,
\begin{equation}
 \kappa = \sum^{2N_r}_{r=1} \left| \frac{\bar{n}_r}{\sigma_n/2} \right|^2 \label{Eq:kappa}
\end{equation}
is a central chi-squared random variable with $2N_r$ degree of freedom having a cumulative distribution function (CDF) equal to~\cite{book:p0001},
\begin{equation}
 F_{\kappa}(a,b) = \frac{\gamma(b/2,a/2)}{\Gamma(b/2)} \label{Eq_50}
\end{equation}
\noindent where $\gamma(c,d)$ is the lower incomplete gamma function given by,
\begin{equation}
 \gamma(c,d) = \int_0^d t^{c-1}e^{-t}dt
\end{equation}
and $\Gamma(c)$ is the gamma function given by,
\begin{equation}
 \Gamma(c) = \int_0^{\infty} t^{c-1}e^{-t}dt
\end{equation}

 The initial sphere radius considered in SM--SD is a function of the noise variance as given in~\cite{xhw0701},
\begin{equation}
 R^2 = 2 \alpha N_r \sigma_n^2 \label{Eq:R}
\end{equation}
\noindent where $\alpha$ is a constant chosen to satisfy \eqref{Eq:NearOP}. This can be done by setting $\Pr\left(\left(\ell_t,s_t\right) \notin \Theta _R\right)=10^{-6}$ and back solving \eqref{Eq:PrNoPoint}. For $N_r = 1,2,4$, $\alpha = 13.8,8.3,5.3$ respectively.

 Finally, $\Pr_{e,\text{SM--SD}}$ can be formulated as,
\begin{eqnarray}
 \Pr_{e,\text{SM--SD}} &=& \Pr\left(\left\| \mathbf{\bar{y}} - \mathbf{\bar{h}_{\ell}}\bar{\bf{s}}\right\|^2 > \left\| \mathbf{\bar{y}} - \bar{\bf{h}}_{\ell_t} \bar{\bf{s}}_t\right\|^2 \right) \nonumber \\
&=& \Pr\left(  \xi > \left\|\mathbf{\bar{h}}_{\ell_t}\bar{\bf{s}}_t-\mathbf{\bar{h}}_{\ell}\bar{\bf{s}}\right\|\right)
\end{eqnarray}
\noindent where,
\begin{equation}
\xi = 2{\rm Re}\left\{\left(\mathbf{\bar{h}}_{\ell_t}\bar{\bf{s}}_t - \mathbf{\bar{h}}_{\ell}\bar{\bf{s}}\right)^T\mathbf{\bar{n}}\right\} \sim \mathcal{N}\left(0,2\sigma_n^2\left(\left\|\mathbf{\bar{h}}_{\ell_t}\bar{\bf{s}}_t - \mathbf{\bar{h}}_{\ell}\bar{\bf{s}}\right\|\right)\right)
\end{equation}

 Thus,
\begin{equation}
\Pr_{e,\text{SM--SD}}= Q\left(\sqrt{\frac{\left\|\mathbf{\bar{h}}_{\ell_t}\bar{\bf{s}}_t - \mathbf{\bar{h}}_{\ell}\bar{\bf{s}}\right\|^2}{2\sigma_n^2}}\right) \label{Eq:Q_1}
\end{equation}
\noindent where $Q(x) = (1/\sqrt{2\pi})\int_{x}^{+\infty}e^{-t^2/2}dt$.

In the case of Rayleigh fading, we can derive the closed form solution for $\mathrm{E}_{\bf{H}}\left\lbrace \Pr_{e,\text{SM--SD}} \right\rbrace$ in \eqref{Eq:BER_1} by employing the solution from~\cite[eq. (62)]{ag9902}. Note that the argument of the $Q$-function in \eqref{Eq:Q_1} can be represented as the summation of $2N_r$ squared Gaussian random variables, with zero mean and variance equal to 1. This means that the argument in the $Q$--function can be described by a central chi--squared distribution with $2N_r$ degrees of freedom.

 The result for $\mathrm{E}_{\bf{H}}\left\lbrace \Pr_{e,\text{SM--SD}} \right\rbrace$ is as given in \eqref{Eq:EFinal},
\setcounter{equation}{46}
where $\sigma_s^2 = \| \bar{\bf{x}}_{\ell_t,s_t} - \bar{\bf{x}}_{\ell,s}\|^2_{\rm F}$ and,

\begin{equation}
 \zeta(c) = \frac{1}{2}\left( 1 - \sqrt{ \frac{c}{1+c} } \right)
\end{equation}

 Plugging \eqref{Eq:EFinal} into \eqref{Eq:BER_1} gives a closed form expression for the BER of SM--SD. In the next section, we show that \eqref{Eq:BER_1} gives a tight approximation of the BER of SM--SD, and that SM--SD offers a near optimum performance.

 \begin{figure}[!b]
    \centering
    \includegraphics[width=9.1cm]{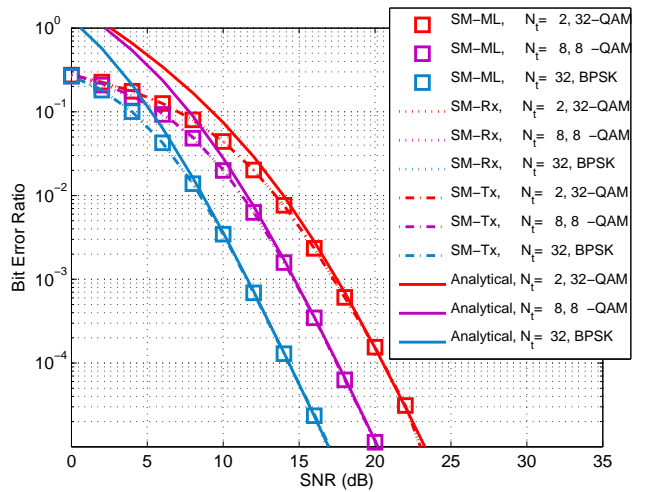}
      \caption{BER against SNR. $m=6$, and $N_r=4$.}
    \label{fig:BER_ANA_SM_6_4}
\end{figure}

\section{Results}\label{sec:Result}

 In the following, Monte Carlo simulation results for at least $10^6$ Rayleigh fading channel realisations are shown to compare the performance and computational complexity of large scale MIMO, SM--ML, SM--SD and SMX--SD.

\subsection{Analytical performance of SM--SD}

\begin{figure}[t!]
    \centering
    \includegraphics[width=8.8cm]{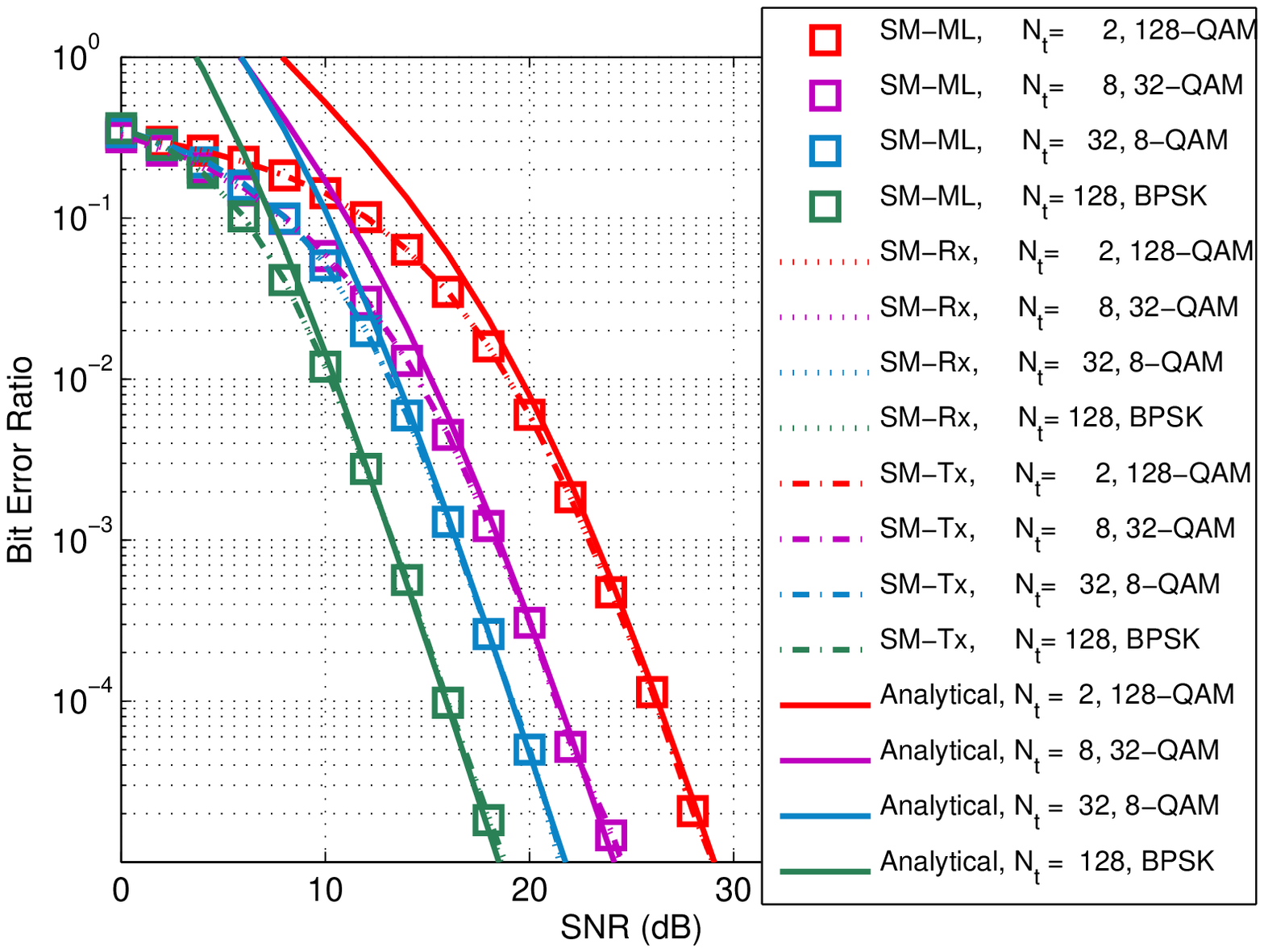}
      \caption{BER against SNR. $m=8$, and $N_r=4$.}
    \label{fig:BER_ANA_SM_8_4}
\end{figure}

 Figs.~\ref{fig:BER_ANA_SM_6_4}-\ref{fig:BER_ANA_SM_8_4} show the BER simulation results for SM--ML, SM--Rx and SM--Tx compared with the analytical bound derived in section~\ref{sec:ANA}, where $m=6,8$ and $N_r=4$. From the figures we can see that both SM--Tx and SM--Rx offer a near optimum performance, where the results overlap with SM--ML. Furthermore, Figs.~\ref{fig:BER_ANA_SM_6_4}-\ref{fig:BER_ANA_SM_8_4} validate our analytical bound as for BER $<10^{-2}$ all graphs closely match the analytical results. Note, it is will--known that the union bound is loose for low SNR~\cite{book:p0001}.

\subsection{Comparison of the BER performance of SM and SMX}

 \begin{figure}[b!]
     \centering
       \includegraphics[width=9.6cm]{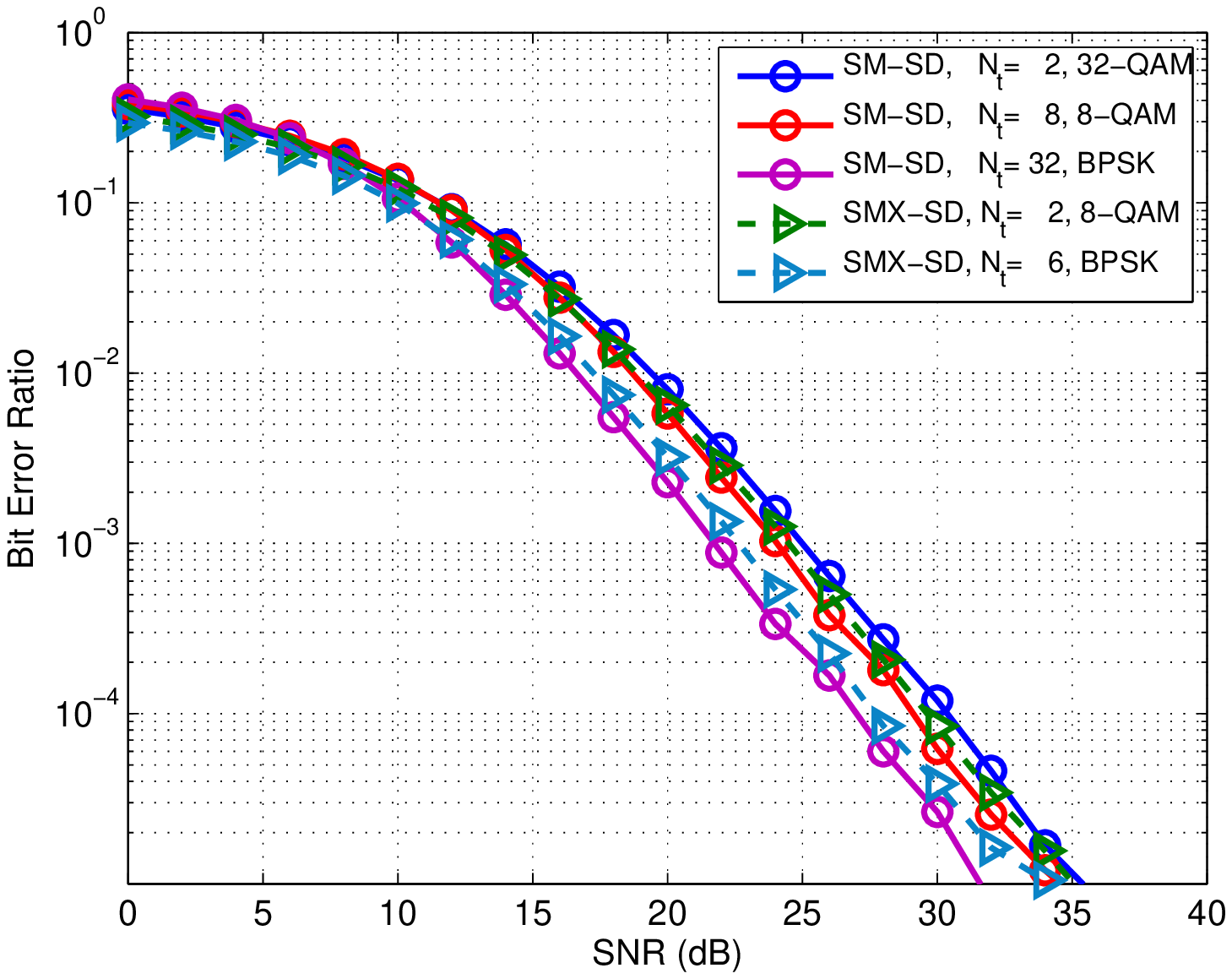}
       \caption{BER against SNR. $m=6$, and $N_r=2$.}
     \label{fig:BER_SM_6_2}
 \end{figure}
 \begin{figure}[htbp]
 \vspace{-0.22cm}
     \centering
       \includegraphics[width=9.6cm]{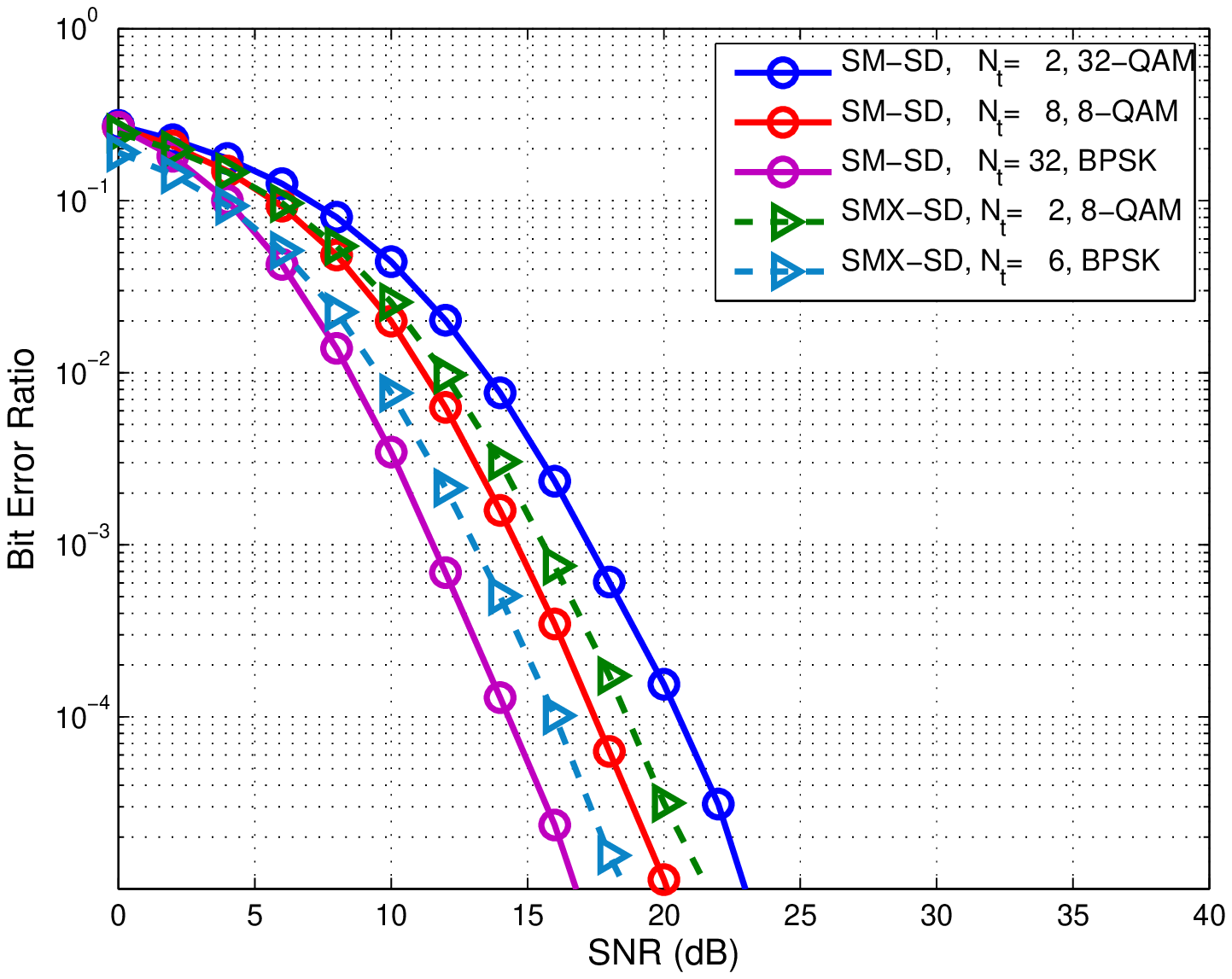}
       \caption{BER against SNR. $m=6$, and $N_r=4$.}
     \label{fig:BER_SM_6_4}
 \end{figure}
 \begin{figure}[htbp]
     \centering
       \includegraphics[width=9.6cm]{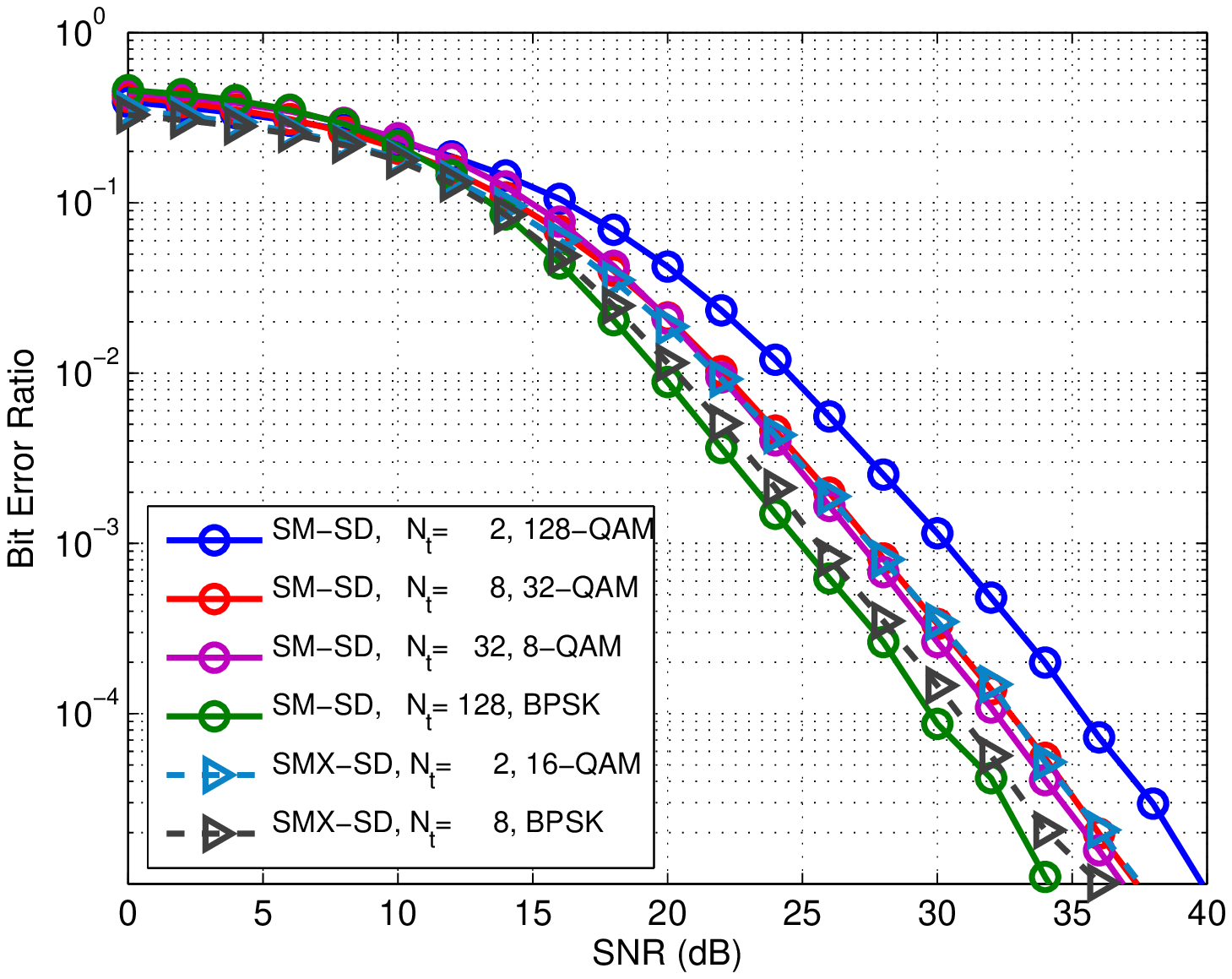}
       \caption{BER against SNR. $m=8$, and $N_r=2$.}
     \label{fig:BER_SM_8_2}
 \end{figure}
 \begin{figure}[htbp]
     \centering
       \includegraphics[width=9.6cm]{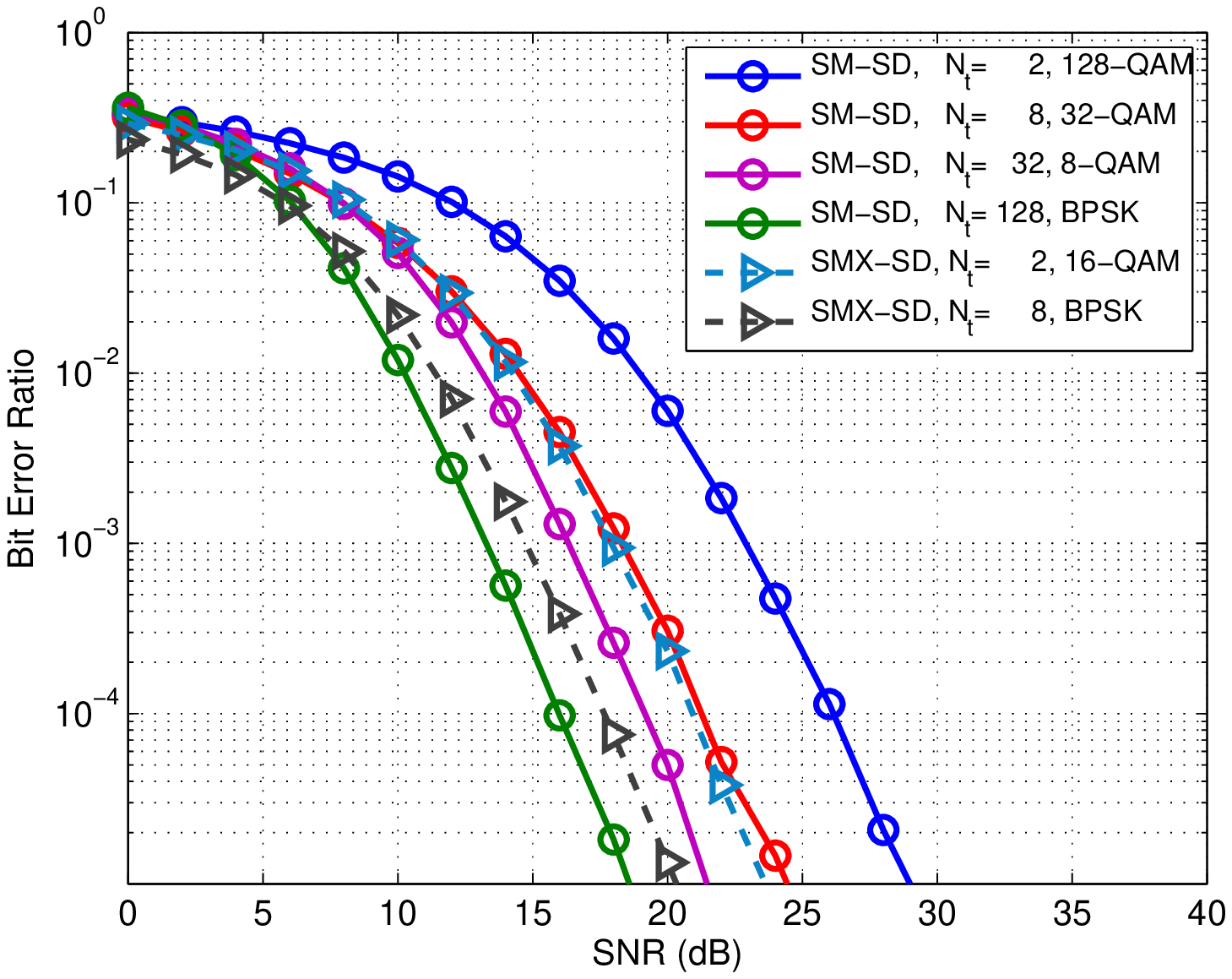}
       \caption{BER against SNR. $m=8$, and $N_r=4$.}
     \label{fig:BER_SM_8_4}
 \end{figure}

\vspace{0.3cm}
 Figs. \ref{fig:BER_SM_6_2} and \ref{fig:BER_SM_6_4} show a BER comparison between all possible combinations of SM and SMX for $m=6$ and $N_r=2,4$. In Fig. \ref{fig:BER_SM_6_2}, we can observe that the BER performance depends on the the number of transmit antennas used and, consequently, the constellation size. The smaller the constellation size, the better the performance. Another observation that can be made is that SM and SMX offer nearly the same performance when using the same constellation size.
 In Fig. \ref{fig:BER_SM_6_4}, where the number of receive antennas is increased, we notice that SM performs better than SMX. In particular, BPSK--SM provides a $1$ dB better performance than BPSK--SMX. Also 8--QAM SM offers a slightly better performance ($\sim 0.5$ dB) than 8-QAM SMX.

 In Figs. \ref{fig:BER_SM_8_2} and \ref{fig:BER_SM_8_4}, the BER comparisons for $m=8$ and $N_r=2,4$ are shown. In Fig. \ref{fig:BER_SM_8_2}, SM and SMX offer similar performance for the same constellation size. However, SM offers a better performance when the number of receive antennas increases as shown in Fig. \ref{fig:BER_SM_8_4}.

  In summary, SM offers a similar or better performance than SMX, where the performance of both systems depends on the size of the constellation diagram and the number of receive antennas. We also note that the BER performance of SM can be improved by increasing the number of receive antennas.

\subsection{Complexity Analysis}

\begin{figure}[!t]
    \centering
      \includegraphics[width=9.6cm]{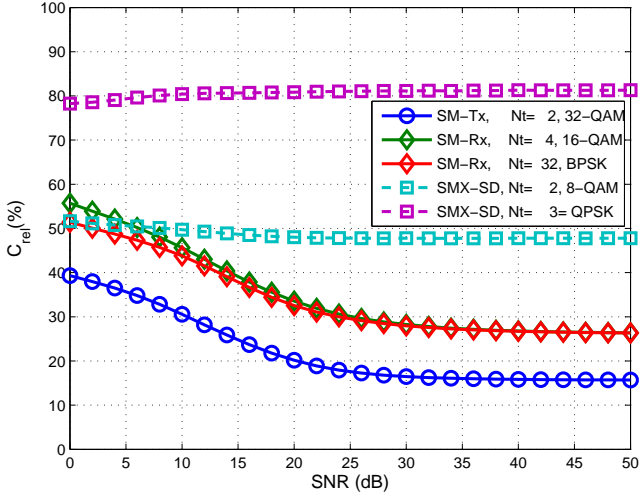}
      \caption{Computational complexity against SNR. $m=6$, and $N_r=2$.}
    \label{fig:Comp_SM_6_2}
\end{figure}
\begin{figure}[!t]
    \vspace{-0.15cm}
    \centering
      \includegraphics[width=9.6cm]{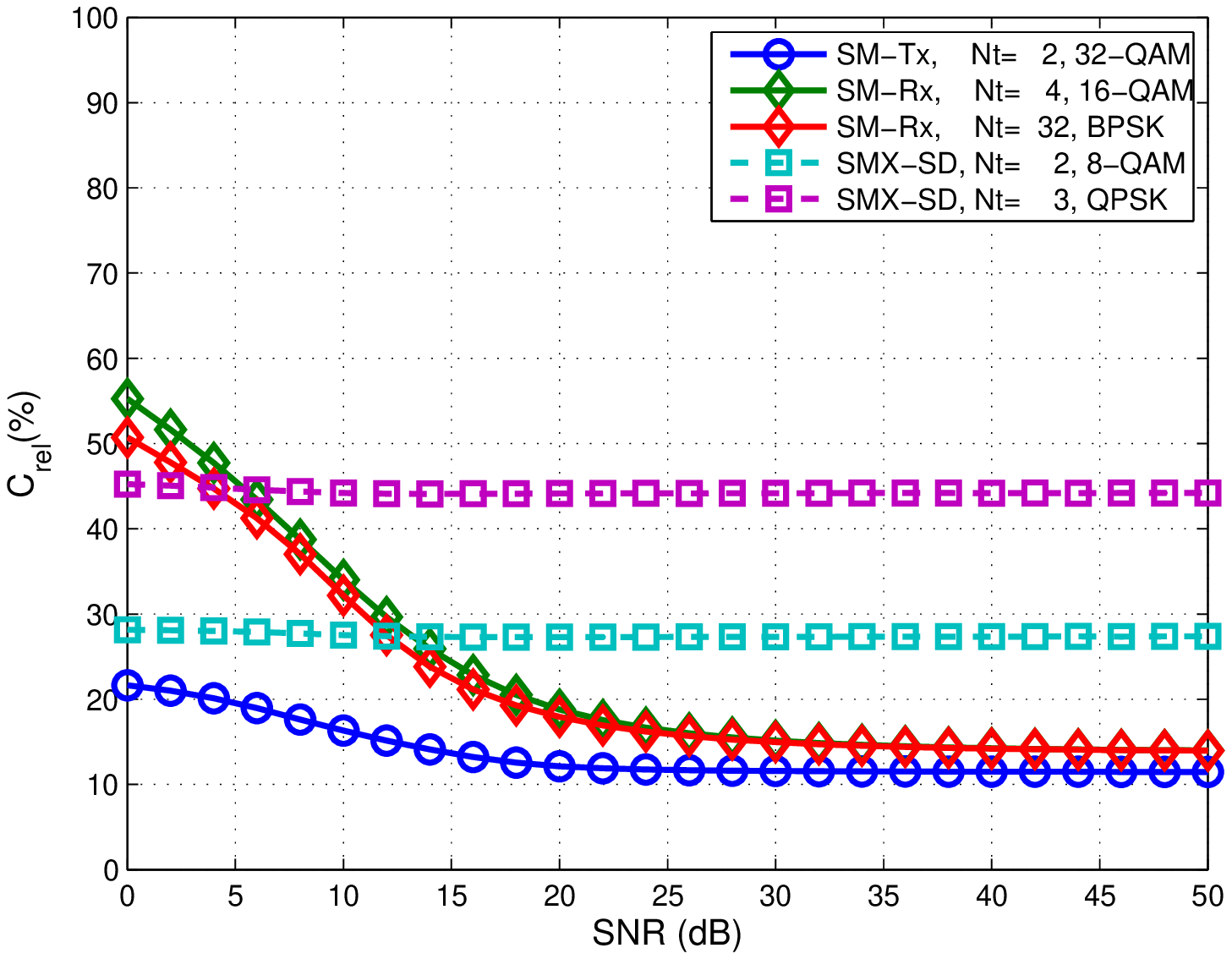}
      \caption{Computational complexity against SNR. $m=6$, and $N_r=4$.}
    \label{fig:Comp_SM_6_4}
\end{figure}

\begin{figure}[!t] 
    \centering
      \includegraphics[width=9.6cm]{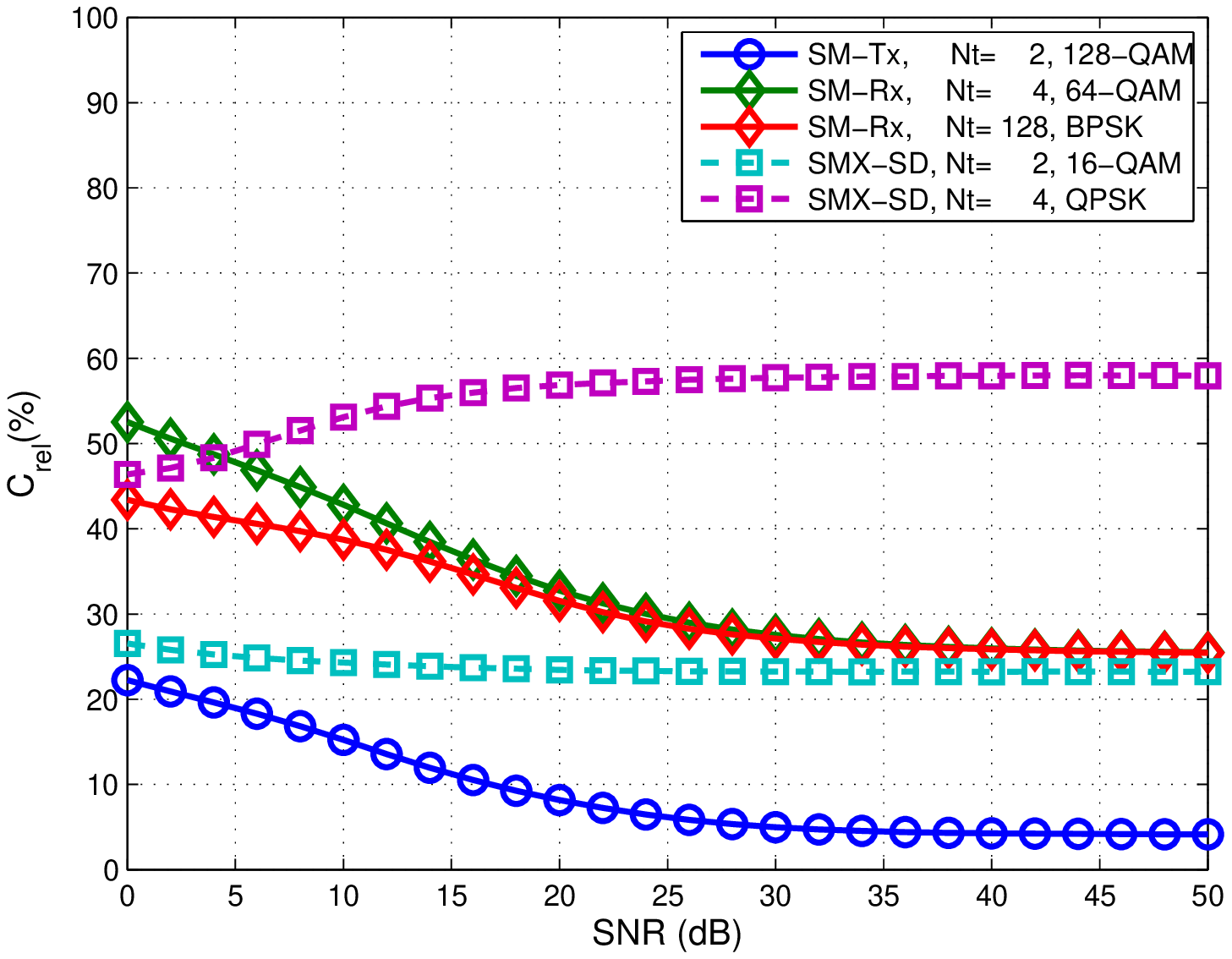}
      \caption{Computational complexity against SNR. $m=8$, and $N_r=2$.}
    \label{fig:Comp_SM_8_2}
\end{figure}
\begin{figure}[!t] 
\vspace{-0.15cm}
    \centering
      \includegraphics[width=9.6cm]{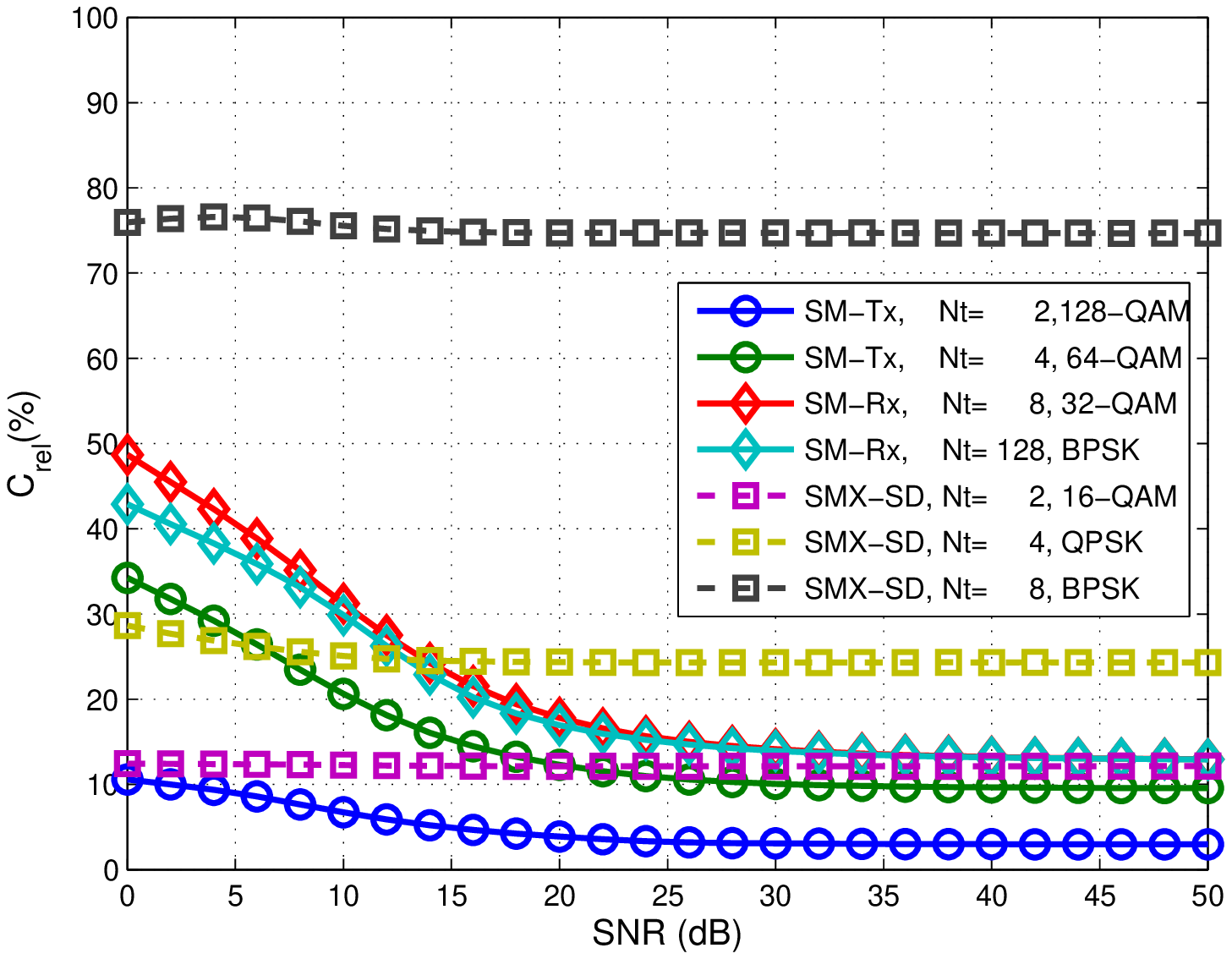}
      \caption{Computational complexity against SNR. $m=8$, and $N_r=4$.}
    \label{fig:Comp_SM_8_4}
\end{figure}

 \begin{figure*}[hb!]
\setcounter{equation}{48}
\normalsize \hrulefill \vspace*{0pt}
\begin{eqnarray}
R^2+\varphi \bar{\bf x}_{\ell,s}^H \bar{\bf x}_{\ell,s} &\geq& \left\lbrace \| \bar{\bf y}-\bar{\bf{H}}\bar{ \bf x}_{\ell,s} \|_{\rm{F}}^2 + \varphi \bar{\bf{x}}_{\ell,s}^H \bar{\bf x}_{\ell,s} \right\rbrace \nonumber \\
&\geq& \left\lbrace \bar{\bf y}^H\bar{\bf y} - \bar{\bf y}^H\bar{\bf H} \bar{\bf x}_{\ell,s} + \bar{\bf x}_{\ell,s}^H\bar{\bf H}^H\bar{\bf y} + \bar{\bf x}_{\ell,s}^H \bar{\bf G}\bar{\bf x}_{\ell,s} \right\rbrace \label{Eq:ineq_2}
\end{eqnarray}
\setcounter{equation}{47}
\end{figure*}

 In Figs. \ref{fig:Comp_SM_6_2}-\ref{fig:Comp_SM_8_4}, the computational complexity of SM--Rx and SM--Tx provided in \eqref{Eq:Rxcomp} and \eqref{Eq:Txcomp} respectively is compared with the computational complexity of SMX--SD, where the initial radius is chosen according to \eqref{Eq:R}. In particular, the figures show the relative computational complexity of the SDs with respect to the SM--ML detector, \emph{i.e}  $\mathcal{C}_{{\rm{rel}}} \left( \%  \right) = 100 \times { \left(\mathcal{C}_{{\rm{SD}}} / {\mathcal{C}_{{\rm{ML}}} }\right)}$. Note, for SM the SD with the lowest complexity is chosen.

 In Figs. \ref{fig:Comp_SM_6_2} and \ref{fig:Comp_SM_6_4}, the relative computational complexities for $m=6$~and $N_r=2,4$ are shown. Fig. \ref{fig:Comp_SM_6_2}, shows that for large constellation sizes the lowest relative computational complexity is offered by SM--Tx $N_t=2$. The relative computational complexity ranges between $40\%$ for low SNR and $16\%$ for high SNR. However, for lower constellation sizes SM--Rx provides the lowest relative computational complexity, which is between $56\%$ for low SNR and $26\%$ for hight SNR. As SM--Rx reduces the receive search space, the reduction in the computational complexity offered by SM--Rx does not depend on the number of transmit antennas. Therefore, only SM--Rx with $N_t=4,32$ are shown, where both scenarios offer nearly the same relative computational complexity. Finally, from Fig. \ref{fig:Comp_SM_6_2} we can see that SMX--SD $N_t=2$ and $N_t=3$ are less complex than SM--ML with a relative computational complexity $48\%$ and $79\%-82\%$ respectively.
 However, comparing SM--SD to SMX--SD $N_t=2$, for $32$--QAM SM--SD is $32\%$ less complex than SMX---SD, and for BPSK SM--SD is $22\%$ less complex than SMX--SD.
 In Fig. \ref{fig:Comp_SM_6_4}, it can be seen that for large constellation sizes SM--Tx is still the best choice with a relative complexity that ranges between $22\%$ for low SNR and $12\%$ for hight SNR, which is $15\%$ less than SMX--SD $N_t=2$. For smaller constellation sizes SM--Rx is the best choice with relative complexity that ranges between $55\%$ for low SNR and $14\%$ for high SNR, offering a $23\%$ extra reduction in complexity when compared to SMX--SD $N_t=2$. Note, i) SMX--SD $N_t=6$ is not shown in the figure, because this scenario has a complexity higher than the complexity of SM--ML, ii) the complexity of SMX--SD $N_t=3$ increased with the increase of SNR, for the reason that, in the under-determined case $\varphi$ depends on the SNR \eqref{eq_varpi}.

 The relative complexity for $m=8$ and $N_r=2,4$ is shown in Fig. \ref{fig:Comp_SM_8_2} and \ref{fig:Comp_SM_8_4}.
 Since SM--Tx reduces the transmit search space, the reduction in complexity increased by more than $10\%$ with the increase in the wordsize and consequently the constellation size. In Fig. \ref{fig:Comp_SM_8_2} for high constellation sizes SM--Tx $N_t=2$ is the best choice with a relative complexity that reaches $4\%$ for high SNR,. In Fig. \ref{fig:Comp_SM_8_4} for high constellation sizes SM--Tx $N_t=2$ and $N_t=4$ are the best choice with a relative complexity that reaches $3\%$ and $10\%$ respectively. On the other hand, SM--Rx reduces the receive search space, therefore, it still offers nearly the same relative complexity. However, the complexity reduces with the increase of $N_r$, where SM--Rx $N_r=4$ is $(\sim 10\%)$ less complex than SM--Rx $N_r=2$.
 Finally, from both figures it can be seen that although SM--ML is much less complex than SMX--ML, SMX--SD is less complex than SM--ML. For that reason, SM--SD has to be developed, where SM--SD is  $(\sim 20\%)$ less complex than SMX--SD for $N_r=2$, and $(\sim 10\%)$ less complex than SMX--SD for $N_r=4$. Note, the complexity of both SM--Tx and SMX--SD decreases with the increase of $N_r$, because for the case of $N_r<N_t$, the less under-determined the system, the fewer pre--computations are needed.

 To summarize, two SDs for SM are introduced: SM--Tx which reduces the transmit search space, and SM--Rx which reduces the receive search space. Both detection algorithms are shown to offer a significant reduction in computational complexity while maintaining a near optimum BER performance. For systems with few transmit antennas, SM--Tx is shown to be the better choice. For systems with with a larger number of receive antennas, SM--Rx is shown to be the better candidate in terms of complexity reduction. The decision for the most appropriate SD depends on the particular deployment scenario.

\section{Conclusion}\label{sec:Con}

 In this paper we have introduced and analysed the performance/complexity trade--off of two SDs designed specifically for SM. The proposed SDs provide a substantial reduction in the computational complexity while retaining the same BER as the ML--optimum detector. The closed--form analytical performance of SM in i.i.d. Rayleigh flat--fading channels has been derived, and analytical and simulation results were shown to closely agree. Furthermore, numerical results have highlighted that no SD is superior to the others, and that the best solution to use depends on the MIMO setup, and the SNR at the receiver. In general, SM--Rx is the best choice for lower spectral efficiencies, and SM--Tx is the best option for higher spectral efficiencies. Finally, simulation results showed that SM using SD offers a significant reduction in computational complexity and nearly the same BER performance as SMX using ML decoder or SD.

  Overall, SM--SD offers i) hardware complexity and power consumption that does not depend on the number of transmit antennas, ii) BER performance that increases with the increase of the number of transmit antennas, and iii) a large reduction in computational complexity compared to SMX. Thus, we believe that SM--SD is an ideal candidate for large scale MIMO systems.

 \appendix[Proof of the intervals \eqref{Eq:int1}, \eqref{Eq:int2}] \label{App:Int}

{\em Proof:}

\begin{enumerate}
 \item First \eqref{Eq:Tx-SD} can be thought of as an inequality,
 
\begin{equation}
R^2\geq \left\lbrace  \| \bar{\bf y}-\bar{\bf{H}}\bar{ \bf x}_{\ell,s} \|_{\rm{F}}^2  \right\rbrace  \label{Eq:ineq_1}
\end{equation}

Then add $\varphi \bar{\bf{x}}_{\ell,s}^H \bar{\bf x}_{\ell,s}$ to both sides of \eqref{Eq:ineq_1} to get \eqref{Eq:ineq_2},
 \setcounter{equation}{49}
where $\bf{\bar{G}}=\bf{\bar{H}^H\bar{H}}+\varphi\bf{\bar{I}}_{N_t}$ is a $(2N_t \times 2N_t)$ positive definite matrix, with a Cholesky factorisation defined as $\bf{\bar{G}}=\bf{\bar{D}^H\bar{D}}$, where $\bf{\bar{D}}$ is a $(2N_t \times 2N_t)$ upper triangular matrix.

 Now by defining $\bar{\rho}=\bf{\bar{G}^{-1}\bar{H}^H\bar{y}}$, and adding $\bar{\rho}\bf{\bar{D^HD}}\bar{\rho}$ to both sides of \eqref{Eq:ineq_2}, it can be re--written as,
 
\begin{eqnarray}
 R_{\varphi}^2 &\geq& \left\lbrace \| \bar{\bf z} - \bar{\bf D}\bar{\bf x}_{\ell,s} \|^2_{\rm{F}} \right\rbrace \nonumber \\
      &\geq& \sum\limits_{i = 1}^{2N_t } {\left( {\bar z_i  - \sum\limits_{j = i}^{2N_t } {\bar D_{i,j}
{\bar{ x}_{\ell,s} \left( j \right)} } } \right)^2 }  \label{Eq_R_deri}
\end{eqnarray}

\noindent where, $\bar{\bf z}=\bar{\bf D}\bar{\rho}$ and,

\begin{eqnarray}
 R_{\varphi}^2 &=& R^2 + \varphi \bf{\bar x^T_{\ell,s} \bar x_{\ell,s}} + \bf{\bar y^T \bar H \bar \rho} -\bf{\bar y^T \bar y} \\
 \varphi &=& \left\{ \begin{array}{rl}
                    \sigma_n^2 & N_t>N_r \\
		    0          & N_t\leq N_r \\
                   \end{array} \right. \label{eq_varpi}
\end{eqnarray}

\noindent For simplicity, in this paper we assume $R_{\varphi}^2=R^2$.

\item Second, we note a necessary condition that the points of the transmit search space need to satisfy to belong to the subset ${\Theta _R }$ is (for all $i=1,2,\ldots,2N_t$):
\begin{equation}
\label{Eq_42} R^2 \geq \left( {\bar z_i  - \sum\limits_{j = i}^{2N_t } {\bar D_{i,j} \bar{ x}_{\ell,s}\left(j \right)} } \right)^2
\end{equation}
\noindent which is a condition similar to conventional SD algorithms~\cite{hv0501a}.
\item We need to take into account that in SM only a single antenna is active at any time instance. In the equivalent real--valued
signal model in \eqref{Eq_9}, this is equivalent to having only two, out of $2N_t$, non--zero entries in the signal vectors
${\bf{\bar x}}_{\ell_t ,s_t }$ and ${\bf{\bar x}}_{\ell ,s }$, respectively. By taking this remark into account, it follows that: a) if
$i=N_t+1,N_t+2,\ldots,2N_t$, then only the imaginary part of ${\bf{\bar x}}_{\ell ,s }$ plays a role in (\ref{Eq_42}), and, thus, only one
entry ${\bar{\bf x}_{\ell,s}(\ell+N_t)}$ can be non--zero; and b) if $i=1,2,\ldots,N_t$, then both real and imaginary parts of ${\bf{\bar x}}_{\ell ,s }$ play a role in (\ref{Eq_42}), and, thus, only two entries ${\bar{\bf x}_{\ell,s}(\ell),\bar{\bf x}_{\ell,s}(\ell+N_t) }$ can be non--zero. The
considerations in a) and b) lead to the intervals in \eqref{Eq:int1} and \eqref{Eq:int2}, respectively, which are directly obtained by solving the
inequality in (\ref{Eq_42}).

\hfill $\Box$
\end{enumerate}

\bibliographystyle{IEEEtran}


\begin{biography}[{\includegraphics[width=1in,height=1.25in,clip,keepaspectratio]{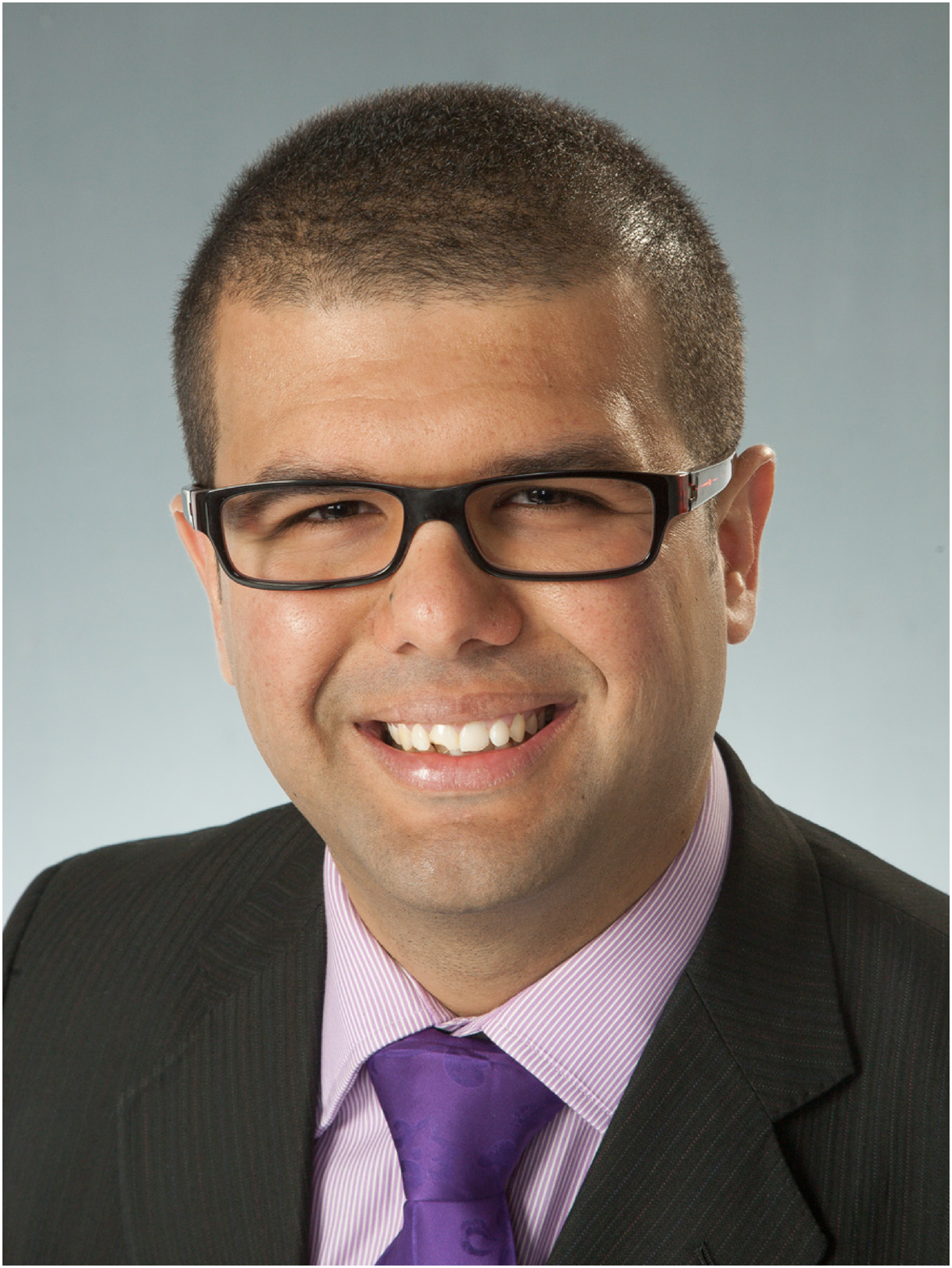}}]{Abdelhamid Younis} received a BSc in Electrical and Electronic Engineering with honours in 2007 from the University of Benghazi, Libya and an MSc with distinction in Signal Processing and Communication Engineering in 2009 from the University of Edinburgh, UK. He is currently completing a Ph.D. in Communication Engineering at the Institute of Digital Communications (IDCOM) at the University of Edinburgh where in 2010 he was awarded the Overseas Research Student Award (ORS) in recognition of his work. His main research interests lie in the area of wireless communication and digital signal processing with a particular focus on spatial modulation, MIMO wireless communications, reduced complexity MIMO design and optical wireless communications.
\end{biography}

\begin{biography}[{\includegraphics[width=1in,height=1.25in,clip,keepaspectratio]{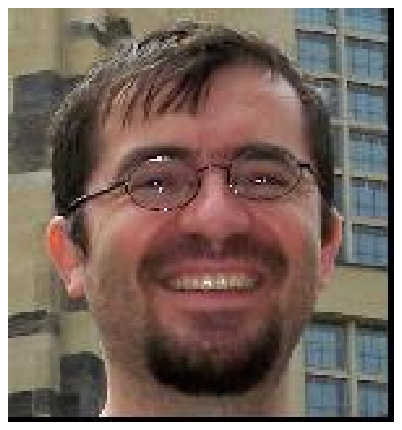}}]{Sinan Sinanovi\'c}
(S'98-M'07) is a lecturer at Glasgow Caledonian University. He has obtained his Ph.D. in electrical and computer engineering from Rice University, Houston, Texas, in 2006. In the same year, he joined Jacobs University Bremen in Germany as a post doctoral fellow. In 2007, he joined the University of Edinburgh in the UK where he has worked as a research fellow in the Institute for Digital Communications. While working with Halliburton Energy Services, he has developed acoustic telemetry receiver which was patented. He has also worked for Texas Instruments on development of ASIC testing. He is a member of the Tau Beta Pi engineering honor society and a member of Eta Kappa Nu electrical engineering honor society. He won an honorable mention at the International Math Olympiad in 1994.
\end{biography}

\begin{biography}[{\includegraphics[width=1in,height=1.25in,clip,keepaspectratio]{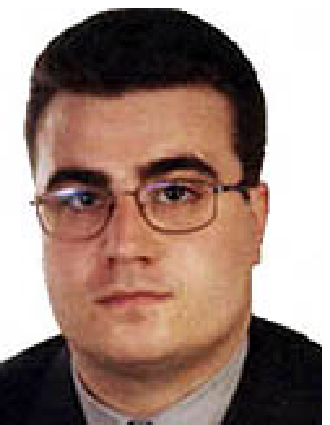}}]{Marco Di Renzo} (S'05--AM'07--M'09) was born in L'Aquila, Italy, in 1978. He received the Laurea (cum laude) and the Ph.D. degrees in Electrical and Information Engineering from the Department of Electrical and Information Engineering, University of L'Aquila, Italy, in April 2003 and in January 2007, respectively.

From August 2002 to January 2008, he was with the Center of Excellence for Research DEWS, University of L'Aquila, Italy.
From February 2008 to April 2009, he was a Research Associate with the Telecommunications Technological Center of Catalonia (CTTC), Barcelona, Spain.
From May 2009 to December 2009, he was an EPSRC Research Fellow with the Institute for Digital Communications (IDCOM), The University of Edinburgh, Edinburgh, United Kingdom (UK).

Since January 2010, he has been a Tenured Researcher (``Charg\'e de Recherche Titulaire'') with the French National Center for Scientific Research (CNRS), as well as a faculty member of the Laboratory of Signals and Systems (L2S), a joint research laboratory of the CNRS, the \'Ecole Sup\'erieure d'\'Electricit\'e (SUP\'ELEC), and the University of Paris--Sud XI, Paris, France. His main research interests are in the area of wireless communications theory. He is a Principal Investigator of three European--funded research projects (Marie Curie ITN--GREENET, Marie Curie IAPP--WSN4QoL, and Marie Curie ITN--CROSSFIRE).

Dr. Di Renzo is the recipient of the special mention for the outstanding five--year (1997--2003) academic career, University of L'Aquila, Italy;
the THALES Communications fellowship for doctoral studies (2003--2006), University of L'Aquila, Italy; the Best Spin--Off Company Award (2004), Abruzzo Region, Italy; the Torres Quevedo award for research on ultra wide band systems and cooperative localization for wireless networks (2008--2009), Ministry of Science and Innovation, Spain; the ``D\'erogation pour l'Encadrement de Th\`ese'' (2010), University of Paris--Sud XI, France; the 2012 IEEE CAMAD Best Paper Award from the IEEE Communications Society; and the 2012 Exemplary Reviewer Award from the IEEE WIRELESS COMMUNICATIONS LETTERS of the IEEE Communications Society. He currently serves as an Editor of the IEEE COMMUNICATIONS LETTERS.
\end{biography}

\begin{biography}[{\includegraphics[width=1in,height=1.25in,clip,keepaspectratio]{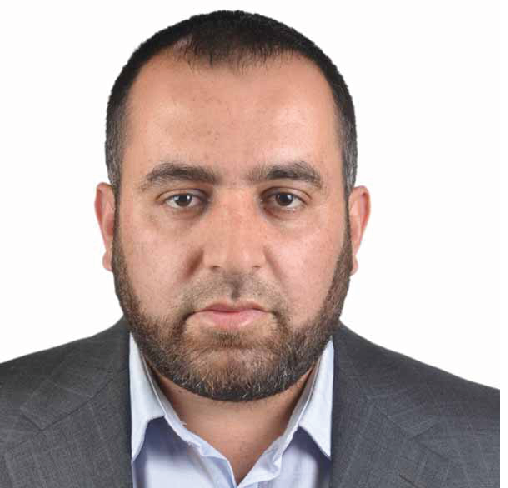}}]
{Raed Mesleh} (S'00-M'08-SM'13) Dr Mesleh holds a Ph.D. in Electrical
Engineering from Jacobs University in Bremen, Germany and several years of
post-doctoral wireless communication and optical wireless communication
research experience in Germany. In October 2010, he joined University of
Tabuk in Saudi Arabia where he is now an assistant professor and the
director of research excellence unit. His main research interests are in
spatial modulation, MIMO cooperative wireless communication techniques and
optical wireless communication. Dr Mesleh publications received more than
800 citations since 2007. He has published more than 50 publications in
top-tier journals and conferences, and he holds 7 granted patents. He also
serves as on the TPC for academic conferences and is a regular reviewer for
most of IEEE/OSA Communication Society's journals and IEEE/OSA Photonics
Society's journals.
\end{biography}

\begin{biography}[{\includegraphics[width=1in,height=1.25in,clip,keepaspectratio]{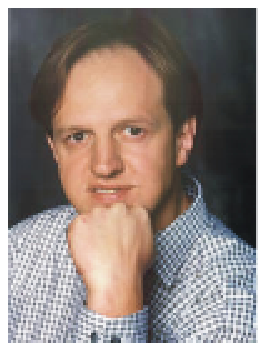}}]
{Professor Harald Haas} (S98-A00-M03) holds the Chair of Mobile
Communications in the Institute for Digital Communications (IDCOM) at
the University of Edinburgh, and he currently is the CTO of a university
spin-out company pureVLC Ltd. His main research interests are in
interference coordination in wireless networks, spatial modulation and
optical wireless communication. Prof. Haas holds 23 patents. He has
published more than 60 journal papers including a Science Article
and more than 160 peer-reviewed conference papers. Nine of his papers
are invited papers. Prof. Haas has co-authored a book entitled ”Next
Generation Mobile Access Technologies: Implementing TDD” with Cambridge
University Press. Since 2007 Prof. Haas has been a Regular High Level
Visiting Scientist supported by the Chinese 111-program at Beijing
University of Posts and Telecommunications (BUPT). He was an invited
speaker at the TED Global conference 2011, and his work on optical
wireless communication was listed among the "50 best inventions in 2011"
in Time Magazine. He recently has been awarded a prestigious Fellowship
of the Engineering and Physical Sciences Research Council (EPSRC) in the UK.
\end{biography}

\end{document}